\PassOptionsToPackage{pdfpagelabels=false}{hyperref} 
\documentclass[a4paper,fleqn,usenatbib,useAMS]{mnras} 

\usepackage{graphicx}	
\usepackage{amssymb}	
\usepackage{multicol}   
\usepackage{bm}			
\usepackage{pdflscape}	
\usepackage{subfig}     
\usepackage[toc,page]{appendix}
\usepackage{hhline}
\usepackage{longtable}
\usepackage{supertabular}
\usepackage{silence} 
\usepackage{makecell}
\usepackage{relsize}
\usepackage[normalem]{ulem}
\usepackage{needspace}
\usepackage{todonotes}	
\usepackage{soul} 

\usepackage{algorithm} 
\usepackage{algpseudocode}

\usepackage{mathtools}

\definecolor{darkgreen}{rgb}{0,0.5,0}

\newcommand{\mcite}[1]{\mbox{\cite{#1}}}
\newcommand{\mcitet}[1]{\mbox{\citet{#1}}}
\newcommand{\mcitealt}[1]{\mbox{\citealt{#1}}}
\newcommand{\mcitep}[1]{\mbox{\citep{#1}}}

\newcommand{\Mpc}{\text{Mpc}}

\newcommand{\assign}[2]{%
  \mathrel{#1}\mathrel{#2}%
}

\newcommand{\llangle}{\left\langle}
\newcommand{\rrangle}{\right\rangle}

\newcommand{\code}[1]{\textsc{\small{#1}}}
\newcommand{\cmfast}{\textsc{\small{21cmFAST}}}
\newcommand{\simfast}{\textsc{\small{SimFast21}}}
\newcommand{\cray}{\textsc{\small{C\textsuperscript{2}-RAY}}}
\newcommand{\pcffast}{\textsc{\small{3PCF-Fast}}}

\renewcommand{\vec}{\mathbf}

\newcommand{\Mmin}{M_{\mathrm{min}}}
\newcommand{\Tvir}{T_{\mathrm{vir}}}
\newcommand{\Rmax}{R_{\mathrm{max}}}
\newcommand{\Rbubble}{R_{\mathrm{bubble}}}
\newcommand{\Zion}{\zeta_{\mathrm{ion}}}
\newcommand{\fcoll}{f_{\mathrm{coll}}}
\newcommand{\deltatb}{\delta T_{\mathrm{b}}}

\newcommand{\xhii}{x_{\mathrm{HII}}}
\newcommand{\xhi}{x_{\mathrm{HI}}}
\newcommand{\meanxhiiz}{\langle x_{\mathrm{HII}}(z) \rangle}

\newcommand{\xip}[1]{\xi^{(#1)}}
\newcommand{\Tspin}{T_{\mathrm{spin}}}
\newcommand{\deltatbr}{\delta T_{\mathrm{b}}(\vec{r})}
\newcommand{\xhiir}{x_{\mathrm{HII}}(\vec{r})}

\newcommand{\eqnref}[1]{Equation~(\ref{#1})}
\newcommand{\secref}{Section~\ref}
\newcommand{\figref}{Figure~\ref}
\newcommand{\tabref}{Table~\ref}

\newcommand{\added}[1]{#1}
\newcommand{\addedagain}[1]{#1}
\newcommand{\removed}[1]{}

\newcommand{\addedtwo}[1]{\textcolor{darkgreen}{\textbf{#1}}}
\newcommand{\removedtwo}[1]{\textcolor{red}{\sout{\substring{#1}{1}{30}}...}}

\renewcommand{\addedtwo}[1]{#1}
\renewcommand{\removedtwo}[1]{}

\usepackage[T1]{fontenc}
\usepackage{ae,aecompl}

\usepackage{amsmath}	

\title[Three-point correlation function to analyse EoR]{Analysing the Epoch of Reionization with three-point correlation functions and machine learning techniques}

\author[W. D. Jennings et al.]{ W. D. Jennings $^{1}$, C. A. Watkinson $^{2, 3}$, F. B. Abdalla $^{1}$
\\
$^{1}$ Department of Physics \& Astronomy, University College London, Gower Street, London WC1E 6BT, UK
\\
$^{2}$ Blackett Laboratory, Imperial College, London, SW7 2AZ, UK
\\
$^{3}$ School of Physics and Astronomy, Queen Mary University of London, Mile End Road, London E1 4NS, UK \\
}

\date{Accepted 2020 August 24. Received 2020 August 13; in original form 2019 July 22}

\pubyear{2020}

\begin{document}
\label{firstpage}
\pagerange{\pageref{firstpage}--\pageref{lastpage}}
\maketitle

\begin{abstract}

Three-point and high-order clustering statistics of the high-redshift 21cm signal contain valuable information about the Epoch of Reionization. We present \pcffast{}, an optimised code for estimating the three-point correlation function of 3D pixelised data such as the outputs from numerical and semi-numerical simulations. \removed{The code includes jackknifing for error estimates and user-changeable parameters for different approximate levels of sampling.}After testing \pcffast{} on data with known analytic three-point correlation function, we use machine learning techniques to recover the \added{mean} bubble size and global ionisation fraction from correlations in the outputs of the publicly available \cmfast{} code. \addedagain{We assume that foregrounds have been perfectly removed and negligible instrumental noise.} Using ionisation fraction data, our best \added{MLP} model recovers the \added{mean} bubble size with a median prediction error of around \added{$10 \%$}, or from the 21cm differential brightness temperature with median prediction error of around \added{$14 \%$}. A further two \added{MLP} models recover the global ionisation fraction with median prediction errors of around $4 \%$ (using ionisation fraction data) or around $16 \%$ (using brightness temperature). Our results indicate that clustering in both the ionisation fraction field and the brightness temperature field encode useful information about the progress of the Epoch of Reionization in a complementary way to other summary statistics. \added{Using clustering would be particularly useful in regimes where high signal-to-noise ratio prevents direct measurement of bubble size statistics. We compare the quality of MLP models using the power spectrum, and find that using the three-point correlation function outperforms the power spectrum at predicting both global ionisation fraction and mean bubble size.}

\end{abstract}



\section{Introduction}
\label{sec:introduction}
A few hundred million years after the Big Bang, the first stars and galaxies began to form \mcitep{Bromm2009}. The radiation emitted from these luminous structures interacted with the surrounding neutral hydrogen and caused it to become ionized. These initially isolated ionized bubbles grew over time. \added{Around} 1 billion years after the Big Bang the  Universe became fully-ionized, see for instance \mcitealt{Becker2006} and \mcitealt{Gunn1965}. The phase shift from a fully neutral to a fully ionized Universe occurred during the so-called Epoch of Reionization (EoR). Many particulars about this process remain unconstrained by current data, predominantly because there are precious few sources of observable radiation during this time. \added{Another way to observe the process of reionization would be to distinguish regions of ionized hydrogen in the neutral background. The most promising probe for this is the 21cm hyperfine transition of hydrogen, which is only observed in neutral hydrogen}. Measurements of the 21cm signal on the sky thus provide a map of which parts of the Universe were neutral. By observing this signal at different redshifts, these maps can be extended into three-dimensional maps of the neutral hydrogen. The size and clustering properties of the ionized hydrogen bubbles change throughout the EoR.

The 21cm signal is much weaker than other foreground sources at the same frequencies. These strong foregrounds make it difficult to extract the actual 21cm signal. Past and ongoing purpose-built experiments such as the Murchison Widefield Array\footnote{http://www.mwatelescope.org/telescope} (MWA, \mcitealt{Tingay2012}), the Low Frequency
Array\footnote{http://www.lofar.org/} (LOFAR, \mcitealt{Patil2017}), and the Precision Array for Probing the Epoch of Reionization\footnote{http://eor.berkeley.edu/} (PAPER, \mcitealt{Ali2015}) have begun to place upper limits on the overall intensity of the signal. The Experiment to Detect the Global EoR Signature\footnote{https://www.haystack.mit.edu/ast/arrays/Edges/} (EDGES) last year claimed a first detection of the 21cm signal. This 21cm absorption profile was observed at redshifts between $15 < z < 20$ with an amplitude of $500 \text{mK}$, published in \mcitet{Bowman2018}. This exciting result has generated much attention in the past year, as the amplitude is significantly \added{more negative} than that anticipated by standard reionization models. The \added{strongly negative} amplitude is difficult to explain without considering additional cooling mechanisms or a higher background radiation than that of the Cosmic Microwave Background. Several recent publications have considered possible modifications that could explain the discrepancy, for instance: considering dark matter interactions (\mcitealt{Fialkov2018}, \mcitealt{Barkana2018a}, \mcitealt{Munoz2018}); the properties of dark matter (\mcitealt{Yoshiura2018}, \mcitealt{Fraser2018}, \mcitealt{Yang2018}, \mcitealt{Lawson2018}); axionic dark matter (\mcitealt{Moroi2018}, \mcitealt{Sikivie2018}, \mcitealt{Lambiase2020}); the effects of radio-wave background (\mcitealt{Ewall-Wice2018}); and considerations of mirror neutrinos (\mcitealt{Sierra2018}). Other attempts have been made to explain the amplitude in terms of the foreground analysis method (e.g. \added{\mcitealt{Sims2019}}). However, until other 21cm observations confirm this detection it is still sensible to continue work with the standard fiducial models that exclude such exotic physics.

Upcoming experiments such as the Hydrogen Epoch of Reionization Array\footnote{http://reionization.org/} (HERA, \mcitealt{DeBoer2016}) and the Square Kilometre Array\footnote{https://www.skatelescope.org/} (SKA, \mcitealt{Mellema2013}) will be able to provide more detailed measurements and should allow us to understand the processes of reionization in detail and confirm the scenarios proposed to explain the EDGES detection. The most detailed theoretical modelling of the 21cm signal currently makes use of simulations. Numerical and semi-numerical simulations encapsulate many aspects of the complex non-linear reioinization processes. Common numerical simulations include \cray{} \mcitep{Mellema2005}, which models the ionising photons emission processes and traces these rays from source to absorption; \added{\code{GRIZZLY} (\mcitealt{Ghara2014})}, which uses one-dimensional radiative transfer simulations to model the radiation profiles around different source types, and then stamps these profiles onto source locations; and many codes which use adaptive refinement \mcitep{Kravtsov1997} to model both large scales and small scales in a single simulation (see for example the Cosmic Reionization on Computers program \mcitealt{Gnedin2014}, and LICORICE \mcitealt{Semelin2007}). Such simulations can provide theoretical predictions for a range of possible reionization scenarios, by specifying different values for simulation input parameters.

Comparisons between 21cm data and theory often make use of fast approximate semi-numerical simulations \added{such as \cmfast{} \mcitep{Mesinger2010} and \simfast{} \mcitep{Santos2010}}. By running a large number of simulations for a range of different reionization scenarios, we can determine which scenarios give rise to the best match between simulated and observed data. Two techniques can make this process more efficient. First, sampling methods such as Markov chain Monte Carlo (MCMC) (\mcitealt{Greig2015}, \mcitealt{Hassan2016},  \mcitealt{Pober2016}, \mcitealt{Greig2016}, \mcitealt{Greig2017}) reduce the total number of simulations that are needed in order to hone-in on the best regions of parameter-space. Second, the simulated and observed data can be compressed before comparing them by using summary statistics. These summary statistics reduce the total size of the data while retaining much of the useful information, and are more robust to modelling and sample variance errors. Common summary statistics are the power spectrum and its higher-order equivalent the bispectrum (\mcitealt{Shimabukuro2016}, \mcitealt{Watkinson2017}, \mcitealt{Majumdar2017}, \mcitealt{Giri2018a}, \mcitealt{Watkinson2018}, \added{and  \mcitealt{Hutter2019}}). Both statistics contain information about the clustering properties of ionized hydrogen bubbles.

In this paper we use machine learning techniques to investigate using the three-point correlation function (3PCF) as another summary statistic for 21cm data. In particular, we determine whether the 3PCF can inform us about the \added{mean} size of ionized bubbles ($R_{\mathrm{bubble}}$) and about the global ionization fraction $\langle \xhii\rangle$. These statistics provide information about the progress of the EoR and encode useful information about different physical scenarios. \added{They also provide a means to reduce the effect of thermal noise, since they are statistical quantities that are averaged over the entire map. Although the 3PCF should be less affected by noise than full 21cm maps, the power spectrum should be even less affected. We compare the relative performances of using either the power spectrum or the 3PCF with our methodology. This indicates whether the 3PCF likely encodes any extra information about bubble size statistics than does the power spectrum.}

As well as recent work using the bispectrum, some research has focussed on using the 3PCF as a tool for investigating the EoR. \mcitet{Gorce2019} use a derived statistic from the 3PCF to concentrate on phase information. \mcitet{Hoffmann2018} investigate whether the 3PCF of 21cm data can be modelled using a local bias model. Their resulting model makes predictions with around $20 \%$ accuracy for large ionized regions at early times, but breaks down for other scenarios.

Machine learning has already been suggested and used for a number of different applications with 21cm data: to emulate power spectrum outputs quickly from \cmfast{} (\mcitealt{Jennings2019}, \mcitealt{Schmit2017}, \mcitealt{Kern2017}), to derive reionization parameters directly from the 21cm power spectrum (\mcitealt{Shimabukuro2017}), and to derive reionization parameters from 21cm images (\mcitealt{Gillet2018}). \added{\mcitet{Jennings2019} also present a mapping between \cmfast{} and \simfast{} providing a proof of concept for mapping between simulations that predict different EoR histories.} 

\added{In this paper, we run a large representative sample of semi-numerical simulations using \cmfast{}. For each simulation, we calculate the 3PCF of the resulting 21cm maps.} We also measure the characteristic reionization features: the global ionization fraction, and the size distribution of the ionized bubbles. We then use machine learning techniques to determine the relationships between the 3PCF measurements and the  characteristic reionization features.

The rest of the paper is split in to the following sections. \secref{sec:3PCF} describes the mathematical concept behind the 3PCF, and a description of the code implementation. We also test the code on data with known \added{analytical 3PCF}. In \secref{sec:ReionizationModels} we describe current physical models of the reionization process. We include a description of the \cmfast{} code and a summary of the range of reionization scenarios considered in this paper. \secref{sec:ML} gives an overview of the machine learning techniques we use, including the search strategy that we use to find the best possible model. We also summarise the methods used to analyse the performance of the resulting models. In the remaining sections we use our data to learn about the characteristic reionization features: the \added{mean} bubble size in \secref{sec:typical_bubble_size_models}, and the global ionization fraction in \secref{sec:global_ionization_fraction_models}. We end the paper in \secref{sec:conclusions} with our conclusions. For Cosmological parameters, we use $\Omega_{\mathrm{M}} = 0.3153$,  $\Omega_{\mathrm{b}} = 0.0493$,  $\Omega_{\mathrm{\Lambda}} = 0.6847$,  $H_0 = 67.36 \text{km s}^{-1}\text{Mpc}^{-1}$, $n_{\mathrm{s}} = 0.9649$, $\sigma_8 = 0.8111$, the latest results using the default \textit{Plik} likelihood from \mcitet{PlanckCollaboration2018}.

\section{Three point correlation calculation}
\label{sec:3PCF}

The three-point correlation function $\xip{3}(\vec{r_1}, \vec{r_2}, \vec{r_3})$ is defined as the ensemble average over triplets of points in real space,

\begin{equation}
\xip{3} \left( \vec{r_1}, \vec{r_2}, \vec{r_3} \right) = \llangle \delta(\vec{r_1}) \delta(\vec{r_2}) \delta(\vec{r_3}) \rrangle
\end{equation}

\noindent The angular brackets denote an ensemble average over a large region of space (or over a large number of universe realisations) to mitigate the effect of statistical fluctuations. \addedtwo{If the signal is translationally invariant then the ensemble average can be replaced by a spatial average. If the signal is also rotationally invariant then the 3PCF depends only on the lengths ($r_1 = |\vec{r_1}$) of the real-space vectors and not on their directions. }These \addedtwo{invariance} assumptions are broken in real observations, in part due to the light cone effect (for example \citealt{Datta2014}) and redshift space distortions (for example \citealt{Majumdar2015}). Note that the three vectors $\vec{r_1}, \vec{r_2}, \vec{r_3}$ connect three points in real space and so have a vector sum of $\vec{0}$, i.e. they form a closed triangle. In practice, $\xip{3}$ measurements are actually made over \textit{configurations} of triangles, by which we mean over sets of unique triangle side-lengths \added{in order to beat down statistical noise}. The three-point correlation function (3PCF) for a single triangle configuration is an average over all triangles with those side lengths $r_1, r_2, r_3$, namely

\begin{equation}
\xip{3} \left( r_1, r_2, r_3 \right) = \Big \langle \delta(\vec{r_1}) \delta(\vec{r_2}) \delta(\vec{r_3}) \Big \rangle _ { (|\vec{r_1}|, |\vec{r_2}|, |\vec{r_3}|)\ =\ (r_1, r_2, r_3) }.
\end{equation}

\subsection{Code Implementation} 
\label{sec:implementation}

Calculating the 3PCF involves placing differently-sized triangles into the data field. The product of the data values at each of the three triangle vertices is summed over a large number of similarly-sized triangles, and an estimate of the 3PCF is built up. The final output of the algorithm is the 3PCF estimates $\xip{3}(\vec{r_i})$ at a discrete set of radius values $\vec{r_i} = (r_1, r_2, r_3)$, corresponding to a discrete set of radius bins. The 3PCF estimate for each radius bin is calculated by using a set of many triangles with similar (but not identical) side lengths. In this section we first describe how to find these sets of triangles, by matching triangles whose side-lengths lie within a given binned range of radii $R_\mathrm{min} \leq r < R_\mathrm{max}$. \added{We also provide pseudocode for our C++ \pcffast{} algorithm (publicly available on GitHub\footnote{https://github.com/wdjennings/3PCF-Fast}).} Finally we discuss how we use the output statistics from the \pcffast{} code to estimate 3PCF values. Our algorithm is similar in nature to other high-order codes (see for instance \citealt{Gaztanaga2004}), although we subsample both the triangle configurations and the number of lattice points and measure the level of approximation needed for robust estimates of the 3PCF.

\subsubsection{\pcffast{} for equilateral triangles}
\label{sec:matching_triangles}
\label{sec:cross_correlation_statistics}

\begin{figure}
\centering
\includegraphics[width=0.8\columnwidth]{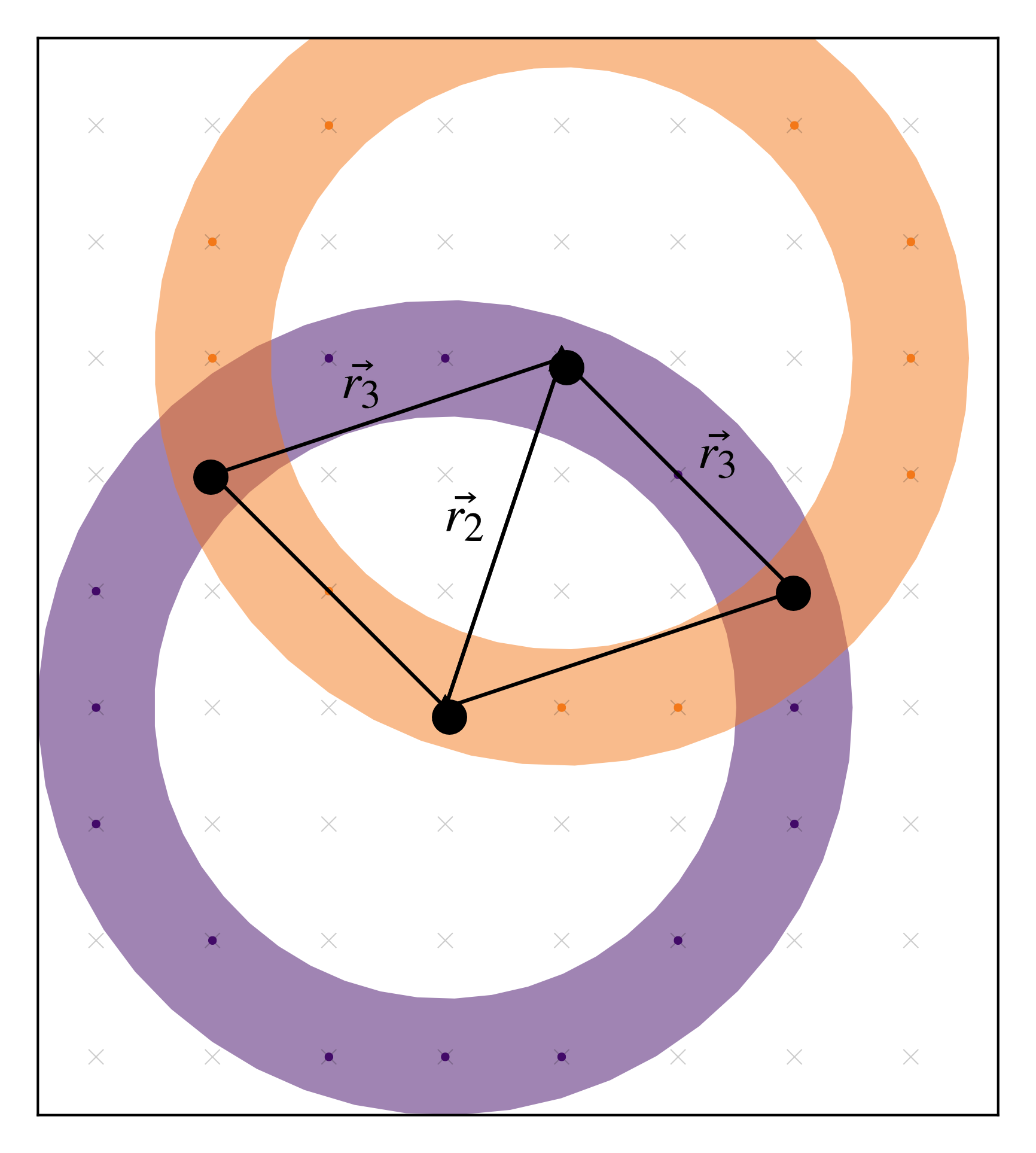}
\caption[Matching triangles to pixels]{Triangles matching the radius bin condition $2.5\ \text{pixels} \leq r < 3.5\ \text{pixels}$. The two regions shown are the radius conditions around the first and second points. The allowed third point(s) then lie in the overlap of these annuli.}
\label{fig:matching_triangles}
\end{figure}

Efficiently finding sets of similarly-sized triangles is a key preparation stage of the algorithm. The data in this section are represented as a pixelised set of scalar values in three-dimensions. For each radius bin we find all the triangles whose edge lengths $r_1, r_2, r_3$ lie within a fixed range of side-lengths $R_\mathrm{min} \leq r_i < R_\mathrm{max}$. There are a finite number of such triangles because the three vertices are constrained to lie on the centres of pixels in the data. To find explicit matching triangles we place the first vertex at the origin. We then find all possible second vertices ($\vec{r_2}$) which lie within the spherical shell $R_\mathrm{min} \leq | \vec{r_2} | < R_\mathrm{max}$ of the origin. From each of the matching second vertex points, we find the third vertex points ($\vec{r_3}$) which are a valid distance both from $\vec{r_2}$ and from the origin. This last step is effectively finding pixels which lie in the overlap of two spherical shells. \figref{fig:matching_triangles} shows an example in two dimensions: with the first triangle vertex at the origin, the dark purple annulus indicates the allowed region for the second vertex between $R_\mathrm{min}$ and $R_\mathrm{max}$. The orange region then shows the allowed region of third vertices from one of the possible second vertices. The final matching triangles (of which there are two) are outlined in black in the figure. To prevent repeated calculations we use a python script to search for these matching triangles and store the resulting pairs of vectors $(\vec{r_2},\vec{r_3})$ in a binary file. This binary file can be loaded by the main C++ algorithm many times. we refer to these binary files as \textit{vertices} files. \added{Measurements of the 3PCF are calculated by looping over possible lattice points $\vec{r_1}$ and summing the contributions for all triangle configurations $(\vec{r_2}, \vec{r_3})$ at that pixel. Both the number of triangle configurations and the number of lattice points are sampled to give faster calculations, and we investigate different levels of sampling in \secref{sec:subsampling_triangle_configurations}.}

The three-point correlation function of a data field is usually calculated in comparison to a random field \addedtwo{without clustering}. The correlation function then quantifies the extent to which the data field is more clustered than the random \addedtwo{unclustered} field. \addedtwo{The purpose of the random field is to create a comparison for the data field. Using a uniform field can thus be seen as a method for counting the number of triangle configuration occurrences.} The outputs from our three-point code are the auto- and cross-correlation statistics between the data (D) and random (R) fields. For the 3PCF these statistics are the data-data-data statistic (DDD), data-data-random (DDR), data-random-random (DRR) and random-random-random (RRR). DDD is the auto-correlation found by multiplying the data field at all three vertices. DDR is the cross-correlation found by multiplying the data at two vertices and the random field at the final vertex; and so on. These statistics will later be combined to give an estimate of the 3PCF. For a scalar data field, the random field should be uniform with mean equal to the data mean.  Instead, it is practically simpler and mathematically identical to normalise the \textit{data} field to have a mean of unity, so that the random field \addedtwo{averaged within in each pixel} is also everywhere unity. This allows our code to skip the correlation calculations for the random field, since the value of \added{RRR} is equal to the known integer count of triangles. Algorithm \ref{algorithm:corr3} shows the pseudocode for our algorithm, taking as inputs a data field \added{$D[\vec{r}]$} and a binned \textit{vertices} file, and outputting the three-point correlation statistics (DDD, DDR, DRR, and RRR) for each radius bin.

\begin{algorithm}[H]
\caption{\textbf{Three-point correlation algorithm}}
\label{algorithm:corr3}
\begin{algorithmic}[1]
\Procedure{3PCF}{$D[\vec{r}], R_\mathrm{min}, R_\mathrm{max}$}
\State DDD, DDR, DRR, RRR $\leftarrow 0$ 
\Comment{Initialise to zero}
\State Load $(\vec{r_2},\vec{r_3})$
\Comment{using ($R_\mathrm{min}$, $R_\mathrm{max}$) \textit{vertices} file}
\For {all $\vec{r_1}$}
\Comment{over all data pixels}
	\For{each $\vec{r_2}$,$\vec{r_3}$ pair}
	\Comment{over matching triangles}
		\State DDD $\assign{+}{=} D[\vec{r_1}] \times D[\vec{r_1}+\vec{r_2}] \times D[\vec{r_1}+\vec{r_3}]$
		\State DDR $\assign{+}{=} D[\vec{r_1}] \times D[\vec{r_1}+\vec{r_2}]$
		\State DRR $\assign{+}{=} D[\vec{r_1}]$
		\State RRR $\assign{+}{=} 1$
	\EndFor
\EndFor
\State \textbf{return} DDD, DDR, DRR, RRR
\EndProcedure
\end{algorithmic}
\end{algorithm}

Algorithm \ref{algorithm:corr3} outputs the correlation statistics (DDD, DDR, DRR, RRR). An estimate of the 3PCF is found by combining these statistics. The simplest such estimator is given by ratios of the data- and random-field auto-correlations,

\begin{equation}
\xip{3} = \frac{DDD - RRR}{RRR}.
\label{eqn:simple_estimator}
\end{equation}

\noindent Another estimator,

\begin{equation}
\label{eqn:LS_estimator}
\xip{3} = \frac{DDD - 3DDR + 3DRR - RRR}{RRR},
\end{equation}

\noindent from \citep{Landy1993} generally leads to less biased results, because it takes account of cross-correlations between the data and random fields which the simple estimator ignores.

\added{The number of triangles found by this matching algorithm can quickly exceed hundreds of thousands for side lengths larger than around ten pixels. Even running the matching algorithm itself for such side lengths can take several days and, more significantly, using such an exhaustive set of triangles in the correlation algorithm would require years of CPU time. An accurate measurement of the 3PCF can be obtained efficiently by subsampling a small number of triangles from all valid matches. In this section we sample 5000 \addedtwo{occurrences of triangle configurations from the total available number of valid matches}. In \secref{sec:subsampling_triangle_configurations} we discuss the effect of sampling and why a value of 5000 was chosen.} 

\subsection{Testing \pcffast{} using points-on-spheres}
\label{sec:shells}
We test our code by generating three-dimensional realisations for a distribution with a known 3PCF. We compare the measured 3PCF from our code to the theoretical form, to get an indication of the regimes in which the code has good accuracy and precision. Our testing distribution consists of three-dimensional realisations made up of a set of points in a box. First, a large set of points are uniformly placed on the surfaces of many identically-sized spheres. The data are then saved to a data file by overlaying a three-dimensional pixelised grid and counting the number of points in each grid: zero for no points, one for a single point, and so on. For all realisations in this section, the data are represented as a box with side length 100 arbitrary units pixelised into $512^3$ pixels. The amplitude and shape of the theoretical 3PCF for these realisations depend on the sphere radius $R$ and the number density of spheres $n_\mathrm{s}$. We describe a scenario as a particular pair of these two parameters. We also use the number of spheres $N_\mathrm{s} = n_\mathrm{s} \times 10^6$, since all realisations in this section have a fixed box size of $100$ arbitrary units in each of the three dimensions. The equilateral 3PCF of points-on-sphere realisations has a closed analytic form \citep{LorneWhiteway}. For a scenario with parameters $n_\mathrm{s}$ and $R$, the 3PCF for equilateral triangles as a function of the triangle side length $r$ is given by

\begin{equation}
\label{eqn:3PCF_POS_theory}
\xip{3}(r; R, n_\mathrm{s}) = 
\begin{dcases}
\frac{1}{16 \pi^3 R^3 n_\mathrm{s}^2 r^2 \sqrt{3 R^2 - r^2}} & \text{  if } r < R \sqrt{3}, \\
0 & \text{ otherwise}\\
\end{dcases}
\end{equation}

\begin{figure}
\centering
\includegraphics[width=\columnwidth]{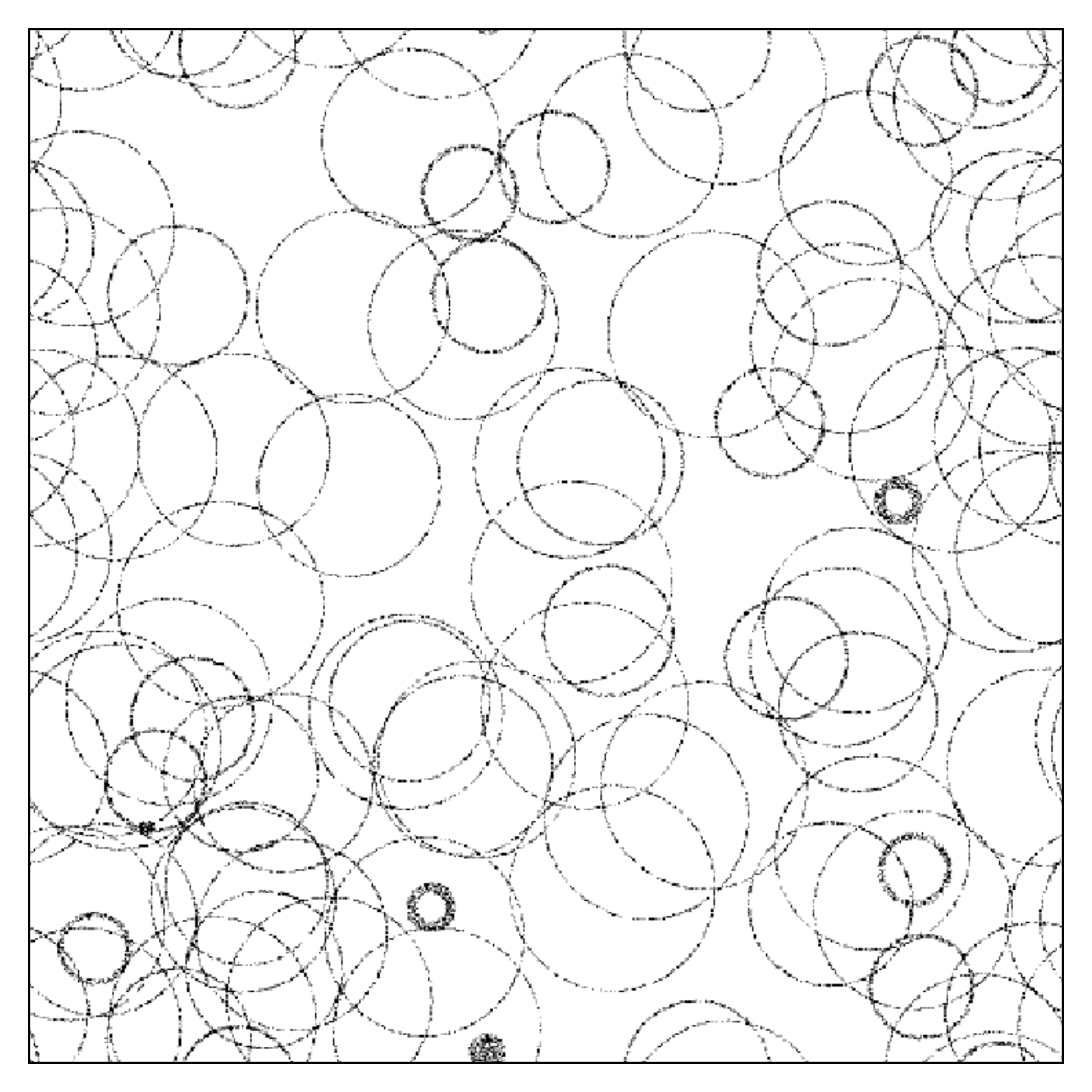}
\caption[Slice through example points-on-spheres realisation]{Slice through an example realisation of points-on-spheres data. This scenario uses spheres with $R=10$ and $N_\mathrm{s}=200$. Each sphere appears as a circular annulus as it has been horizontally sliced for this figure. Some annuli appear thicker than others because slicing a thick spherical shell near its pole gives a wider region when viewed from above.}
\centering
\label{fig:example_data_POS}
\end{figure}

Generating a realisation for a scenario involves choosing where to put the spheres and then placing points on the surfaces of those spheres. A uniformly random set of $N_\mathrm{S}$ points are chosen to be the centres of the spheres. Points are then placed randomly onto the surface of each sphere. Ensuring that the points are indeed uniformly distributed across the spheres surface is most easily done using the method from \citet{Muller1959}: sample three random variables $x,y,z$ from the normal distribution $\mathcal{N}(0,1)$ and divide by the Euclidean norm of these three coordinates. The distribution of the normalised vector

\begin{equation}
\vec{r} = \frac{R}{\sqrt{x^2 + y^2 + z^2}} 
\begin{bmatrix}
x \\ y \\ z
\end{bmatrix},
\end{equation}

\noindent is then uniform across the surface of a sphere with radius $R$. After storing the locations of all points on all spheres, the final pixelised realisation of the scalar field is generated by rounding the point coordinates to the nearest integer. \figref{fig:example_data_POS} shows a slice through an example realisation of the testing distribution. All data in this testing section have pixel size of around $0.2$ arbitrary physical units.

\subsubsection{Test results}
\begin{figure}
\centering
\includegraphics[width=\columnwidth]{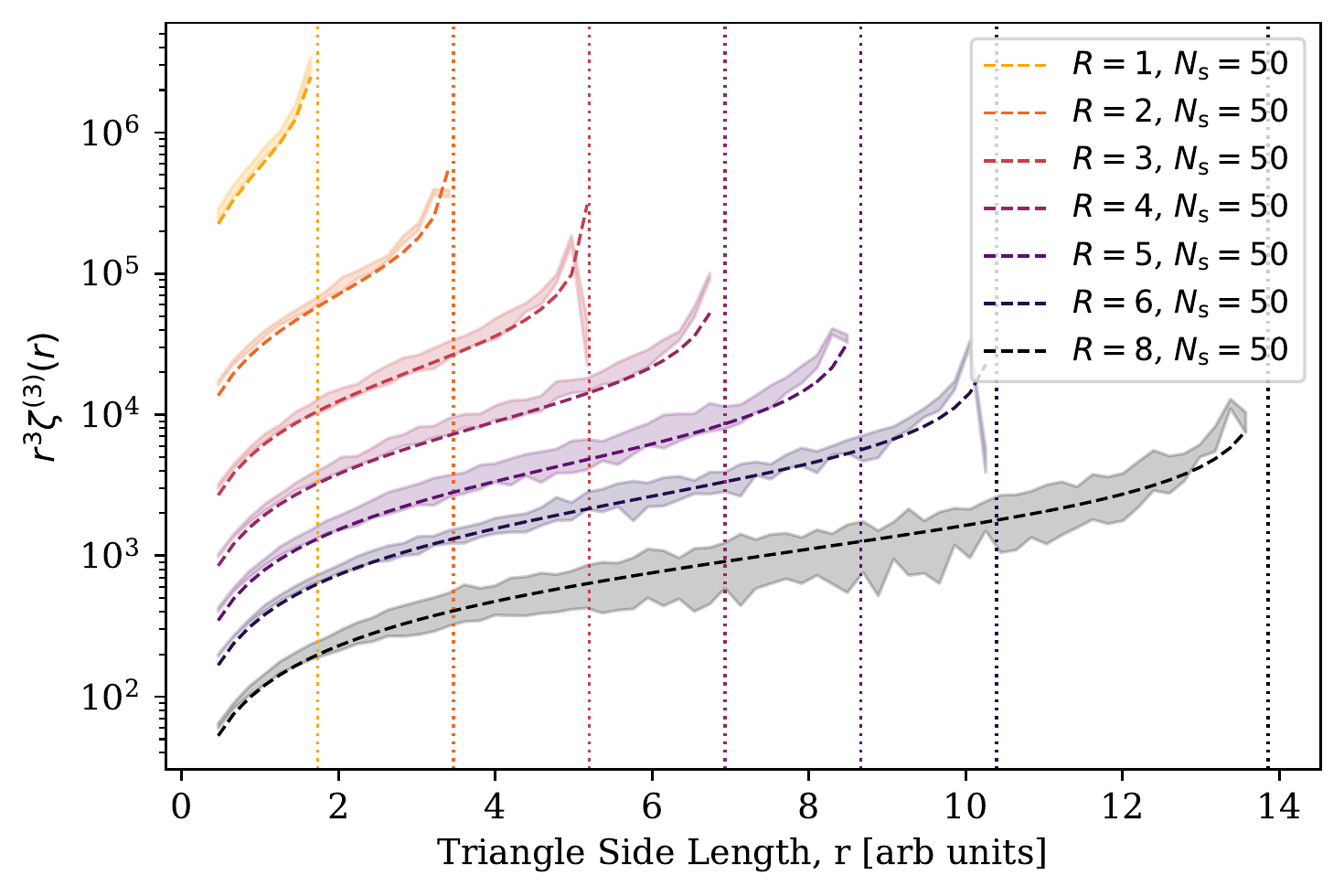}
\caption[Equilateral 3PCF for points-on-spheres data for $R$ scenarios]{Equilateral 3PCFs for points-on-spheres scenarios with varying sphere radius $R$. The dashed lines show the theoretical 3PCF for each scenario whose parameters are shown in the legend. The shaded regions show the standard deviation of the measured 3PCF across five realisations, using our code with the Landy-Szalay estimator. The theory lines lie within the shaded regions for most triangle side lengths, except at radius values near the upper valid limits. Vertical dotted lines show the theory asymptotes at $R\sqrt{3}$ for each line.}
\label{fig:corr3_measured_POS_Rs}
\end{figure}

\begin{figure}
\centering
\includegraphics[width=\columnwidth]{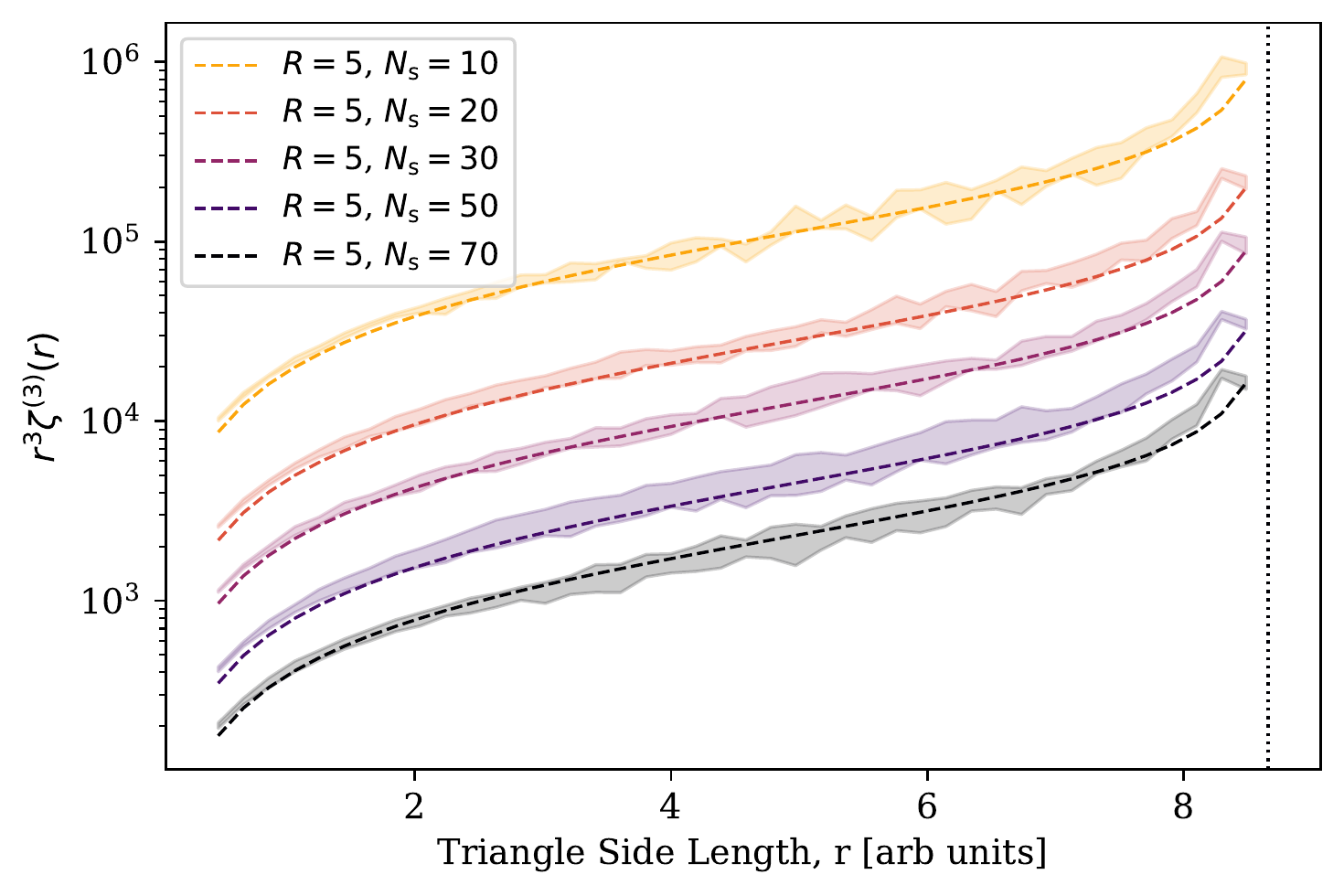}
\caption[Equilateral 3PCF for points-on-spheres data for $N_\mathrm{s}$ scenarios]{Equilateral 3PCFs for points-on-spheres data with varying $n_\mathrm{s}$ between $1\times 10^{-5}$ and $7\times 10^{-5}$, using the LS estimator. The dashed theory lines again lie within the measured shaded regions for most triangle side lengths. Vertical dotted line shows the theory asymptote at $R\sqrt{3} = 5\sqrt{3}$.}
\centering
\label{fig:corr3_measured_POS_Nss}
\end{figure}

We test our code by generating points-on-spheres realisations for many $R$ and $N_\mathrm{s}$ scenarios. We compare the outputs of our code to that of the true theoretical 3PCF using \eqnref{eqn:3PCF_POS_theory}. \figref{fig:corr3_measured_POS_Rs} shows the theoretical and measured equilateral 3PCFs for seven scenarios with a range of $R$ values and fixed $n_\mathrm{s} = 5 \times 10^{-5}$, using the Landy-Szalay estimator in \eqnref{eqn:LS_estimator}. We plot the dimensionless 3PCF defined as $r^3 \xip{3}$ (see for example \added{\citealt{Hoffmann2018}}). The theoretical 3PCF is shown in each case as the dashed line. The measured 3PCF estimates are subject to sample variance, meaning that the output from the code depends on the randomly-seeded initial conditions. We use five realisations with different random seeds to determine whether the theoretical 3PCF lies inside the spread of the five measured code outputs. The shaded regions in \figref{fig:corr3_measured_POS_Rs} show the standard deviation of the measured 3PCF across these five realisations. \figref{fig:corr3_measured_POS_Nss} similarly shows the theoretical and measured 3PCFs for scenarios with fixed $R=5$ and various $N_\mathrm{s}$ values.

The measured and theoretical 3PCFs match closely across most of the triangle side lengths. The theoretical 3PCF in \eqnref{eqn:3PCF_POS_theory} has a vertical asymptote at the maximum allowed radius $R \sqrt{3}$. This can be seen in \figref{fig:corr3_measured_POS_Rs} as a slight upturn near the right-hand sides of each dashed line. Our code slightly over-predicts the theory in each case near the maximum valid radius. This is due to the binning of triangles: each binned output is calculated using equilateral triangles with a range of side lengths as described in \secref{sec:matching_triangles}. Averaging the 3PCF over these differently-sized triangles (some of which are larger than the valid maximum radius) causes a discrepancy between measured and theoretical 3PCF.

\subsection{Optimisation}
\label{sec:subsampling_triangle_configurations}

\begin{figure}
\centering
\includegraphics[width=\columnwidth]{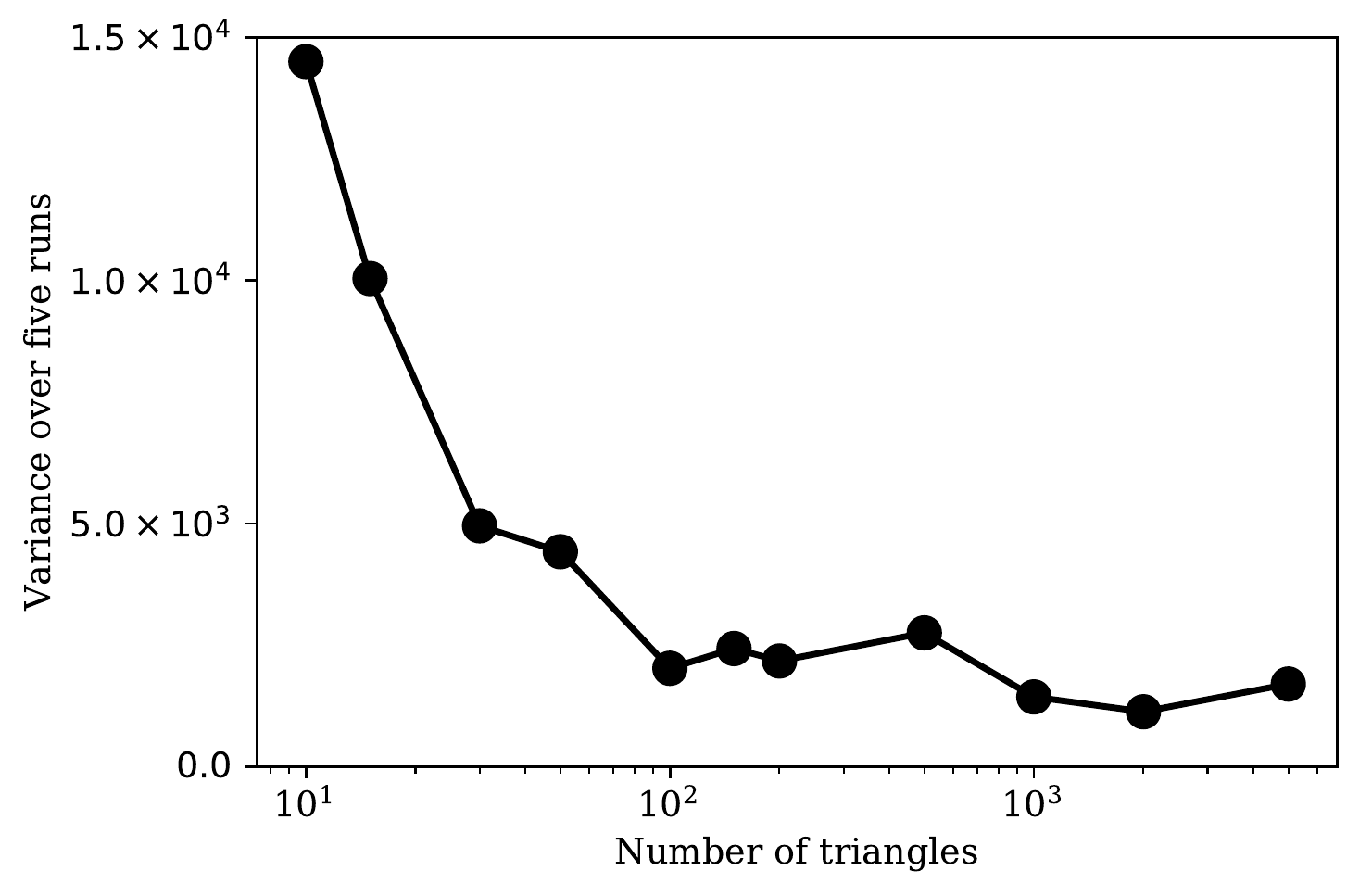}
\caption{Effect of subsampling triangles in the 3PCF algorithm. When few triangles are used, the outputs from the code show a larger variance than when more triangles are used. For more than around $2000$ triangles the variance plateaus indicating that adding more triangles provides minimal extra information.}
\centering
\label{fig:corr3_Ntris}
\end{figure}

A number of steps were taken to optimise and improve the code. First, we added multi-threading \added{to make better use of available computational resources}.  Second, we allow for subsampling of triangle configurations and jackknifing to allow for calculation of errors. The number of triangles found by the matching algorithm in \secref{sec:matching_triangles} can quickly exceed hundreds of thousands for side lengths larger than around ten pixels. \removed{Even running the matching algorithm itself for such side lengths can take several days and, more significantly, using such an exhaustive set of triangles in the correlation algorithm would require years of CPU time.} An accurate measurement of the 3PCF can be obtained more efficiently by subsampling a small number of triangles from all valid matches. We test how this subsampling affects the robustness of the final 3PCF estimate. The 3PCF is calculated on $\xhiir$ data from five randomly-seeded \cmfast{} realisations using input the canonical input parameters $\Tvir = 10^{4} \text{K}$, $\Zion = 30.0$, $\Rmax = 15.0 \Mpc$ and $E_0 = 200\ \text{eV}$. The variance in our code's output over the five realisations is plotted in \figref{fig:corr3_Ntris}, as a function of the number of triangles used in the 3PCF algorithm. The variance is large when only a few triangles are used but decreases with a larger number of triangles. For more than around $2000$ triangles, the scatter plateaus indicating that adding more triangles is unlikely to result in improve final 3PCF estimates. \added{The remaining variance across the five runs is likely due to sample variance.} We use a conservative value of $5000$ triangles in all the 3PCF estimates from hereafter.

\subsubsection{\removed{Estimating sample variance with jackknifing}}
\removed{In the previous subsections we estimate the variance of the 3PCF by taking the spread from differently-seeded random realisations}


\section{Models of reionization}
\label{sec:ReionizationModels}

The 21cm differential brightness temperature $\deltatb$ is defined as the difference between the measured 21cm brightness temperature and the uniform background CMB brightness temperature. By removing the background CMB temperature, the value of $\deltatb (\vec{r})$ then specifies the extent of 21cm emission ($\deltatb > 0$) or absorption ($\deltatb < 0$). The actual observable for radio interferometers is $\deltatb - \langle \deltatb \rangle$, where $\langle \deltatb \rangle$ is the global reionization signal averaged across the whole sky. \mcite{Furlanetto2006} gives an approximate relationship for the 21cm brightness temperature $\deltatb(\vec{r})$ as

\begin{align}
\label{eqn:deltaTb approx}
\deltatb(\vec{r}) = 27 
& x_{\mathrm{HI}}(\vec{r})\ 
\big[ 1 + \delta(\vec{r}) \big]
\left(\frac{\Omega_{\mathrm{b}} h^2}{0.023}\right)
\left(\frac{0.15}{\Omega_{\mathrm{M}} h^2}\right)^{1/2} \\ 
& \left(1 - \frac{T_{\mathrm{\gamma}}}{\Tspin}\right)
\left(\frac{1+z}{10}\right)^{1/2}
\notag \left(\frac{H(z)}{H(z) + \delta_r v_r (\vec{r})}\right)
\ \text{mK}\,.
\end{align}

\noindent This approximation includes the effects of neutral hydrogen fraction $x_{\mathrm{HI}}(\vec{r})$; total matter density contrast $\delta(\vec{r})$; cosmological parameters for the densities of baryonic matter $\Omega_{\mathrm{b}}$ and total matter $\Omega_{\mathrm{M}}$; the CMB temperature $T_{\mathrm{\gamma}}$; the spin temperature $T_{\mathrm{S}}$ which quantifies the relative populations of electrons in the higher and lower energy states of the 21cm transition; the Hubble parameter $H(z)$; and $\delta_r v_r (\vec{r})$, the radial velocity gradient.

The spin temperature can be written \citep{Furlanetto2006} as a sum of three parts,

\begin{equation}
	\Tspin ^{-1} = \frac{T_{\gamma} + x_\alpha T_\alpha^{-1} + x_\mathrm{c} T_\mathrm{K}^{-1}}{1 + x_\alpha + x_\mathrm{c}  }
\end{equation}

\noindent with $T_{\gamma}$ the background CMB temperature, $T_\mathrm{K}$ the kinetic gas temperature, $T_\alpha$ the Lyman-alpha colour temperature (closely linked to the gas temperature for all redshifts of interest), and the coupling coefficients for collisions ($x_\mathrm{c}$) and the Wouthysen-Field coefficient ($x_\alpha$) from \cite{Wouthuysen1952}. In particular, the kinetic gas temperature and the Lyman-alpha background radiation have a strong effect on the global and local evolution of the spin temperature. These two features change throughout the EoR as different physical processes interact with the growth of structure in the Universe.

\subsection{\texorpdfstring{\cmfast{}}{21cmFAST}}
\label{sec:21cmfast}
We use the publicly available semi-numerical code \cmfast{} to generate our data. We briefly describe the algorithm in this subsection. The simulation begins by seeding an initial linear density field onto a three-dimensional grid at very high redshift. This linear density field is evolved using first-order perturbation theory (see \mcitealt{Zeldovich1970}) \addedtwo{to approximate gravitational collapse}, giving \removedtwo{a non-linear } \addedtwo{an approximate gravitationally-evolved} density field $\delta(\vec{r})$.

The simulation then finds the highest density regions where the matter will collapse to form luminous structures and thus contribute ionizing photons towards the reionization process. The extent of collapse is calculated directly from the non-linear density field following the model of spherical collapse\footnote{In order to match the more accurate ellipsoidal collapse model \citep{Sheth1999}, \cmfast{} afterwards normalises the spherical collapse fractions so that their average value matches that expected from ellipsoidal collapse.} \citep{Press1974}. If the mean enclosed density in a region exceeds a theoretical critical value then the region is assumed to collapse. The collapse fraction $\fcoll(\vec{r}, R)$ on decreasing scales $R$ is then found from the contributions of both resolved and unresolved halos. The default \cmfast{} implementation has a minimum halo mass $\Mmin$ for a halo to host star-forming galaxies that evolves with redshift, corresponding to a minimum virial temperature $\Tvir$ for ionising photons. 

The ionization fraction field $x_{\mathrm{HII}}(\vec{r})$ is found by determining whether the collapsed matter in a region generates enough ionizing photons to ionize the enclosed hydrogen atoms. An ionizing efficiency parameter $\Zion$ specifies how many ionizing photons are sourced per unit of collapsed matter. If $\fcoll(\vec{r}, R) \geq \Zion^{-1}$ for any particular region, then the central pixel is painted as fully ionized using the method in \citet{Zahn2007}. this differs from the default method of another common semi-numerical simulation \simfast{}, which instead paints the full spherical region as ionized if there are enough photons using the method in \citep{Mesinger2007}. See \citet{Hutter2018} for a discussion of these two methods. 

Fluctuations in the spin temperature are calculated by considering the kinetic gas temperature and the Lyman-$\alpha$ background temperature. The kinetic gas temperature $T_\mathrm{K}$ is determined by considering the balance between a number of important heating and cooling mechanisms including X-ray emissions, Hubble expansion, adiabatic heating and cooling, and gas particle density changes due to ionization events. The dominant heating effect in \cmfast{} is from X-rays. The rate of emitted X-ray photons is assumed to be proportional to the growth rate of collapsed matter in the dark matter halos. Photons are emitted with a range of wavelengths, the luminosities for which are assumed to follow a power-law relationship $L(\nu) \propto (\nu / \nu_0)^{-\alpha}$. The parameter $\alpha$ controls the slope of this spectral energy density function, and the parameter $\nu_0$ controls the minimum frequency of X-rays which can escape into the Inter-Galactic Medium (IGM). This minimum frequency can also be written in terms of a minimum energy value, $E_0 = h \nu_0$, using the Planck constant\removedtwo{$h = 4.135 \times 10^{-15} \text{eV} \text{s}$}. See \citet{Mesinger2010} for a full derivation of the calculations and assumptions that \cmfast{} makes for the spin temperature fluctuations. 

The final step is to use \eqnref{eqn:deltaTb approx} and calculate the 21cm brightness temperature field $\deltatb (\vec{r})$ using the non-linear density field $\delta(\vec{r})$, the neutral fraction field $x_{\mathrm{HI}}(\vec{r}) = 1 - x_{\mathrm{HII}}(\vec{r})$ and the spin temperature fluctuation field $\Tspin(\vec{r})$.

In this paper we consider different reionization scenarios by changing three of these simulation parameters:

\begin{enumerate}
\item The ionization efficiency $\Zion$, specifying how many ionising photons are sourced per unit of collapsed matter;
\item The $E_0$ parameter which controls the minimum energy (or frequency) of X-ray photons which are able to escape into the IGM;
\item The minimum virial temperature $\Tvir$ which specifies a lower mass limit $\Mmin$ of collapsed matter which produces ionizing photons and X-rays.
\end{enumerate}  

Fixing the other simulation parameters involves setting the efficiency of X-rays to a constant value. We use $\zeta_{\mathrm{X}} = 10^{-57} M_{\odot}^{-1}$ to match the assumption in \citet{Mesinger2010}, equivalent to approximately a single X-ray photon for each stellar baryon as motivated by observations of low-redshift galaxies. The uncertain intergalactic-medium X-ray properties are then parametrised by $E_0$.

\subsection{Training and testing data details}
\label{sec:sample_parameter_space}

We run 1000 \cmfast{} simulations in total for our data. Each simulation generates three-dimensional realisations of the $\deltatb$ field in a cube of size $250\text{Mpc}$ resolved into $256^3$ pixels (smoothed from density fields resolved into $768^3$ pixels). \added{Each simulation uses a different random seed for the initial conditions}. \removed{As discussed in} The resulting redshifts from this algorithm are between $z=5$ and $z=26.6$ (see \cite{Mesinger2010} for a description of the iterative algorithm that generates these steps). These redshifts are 5.0, 5.6, 6.3, 7.0, 7.78, 8.7, 9.6, 10.7, 11.9, 13.2, 14.6, 16.1, and 17.8. \added{We ignore simulated results for higher redshifts, because the mean ionization fraction is extremely small and the mean bubble size is generally smaller than the resolution of our simulations.} For each simulation, we calculate the statistics of interest: the 3PCF using \pcffast{} described in \secref{sec:3PCF}; the bubble size distribution, described in this section; and the global ionization fraction, found by trivially averaging the ionization fraction field $\xhiir$ for each redshift.

In order to sample a range of different reionization scenarios, we use a Latin Hypercube \mcitep{McKay1979} approach\footnote{Using the implementation from \citet{Agarwal2013}}. This method efficiently samples the input space with far fewer simulations than a naive exhaustive grid-search would require. The following ranges and scales of simulation parameters are used:

\begin{enumerate}
\item $\Tvir$ in the logarithmic range $[10^4, 2 \times 10^5]\ \text{K}$
\item $\Zion$ in the linear range $[5, 100]$
\item $E_0$ in the linear range $[100, 1500]\ \text{eV}$
\end{enumerate}

\noindent These ranges were chosen to match those by the simulation authors (for example \mcitealt{Greig2015}). The lower $\Tvir$ limit comes from a minimum temperature for the cooling of atomic hydrogen accreting onto halos. The upper limit arises from observations of high-redshift Lyman break galaxies \citep{Greig2015}. The $\Zion$ upper and lower limits correspond \addedtwo{roughly} to escape fractions of 5\% to 100\% for ionizing photons\addedtwo{ for standard values of the other controlling factors in \citet{Greig2015} such as the number of ionising photons produced per stellar baryon}. The range for $E_0$ was chosen in a similar way to \citet{Park2018}, motivated by hydrodynamic simulations \citep{Das2017} and considering the energy that would allow an X-ray photon to travel a distance of roughly one Hubble length when travelling through a medium with $\meanxhiiz = 0.5$.

\subsubsection{Three-point correlation-function measurements}
\label{sec:measure_corr3}

We use the code described in \secref{sec:3PCF} to calculate the three-point correlation function. We calculate $\xip{3}$ of both the ionization fraction field $\xhiir$ and of the 21cm differential brightness temperature field $\deltatbr$. In the code we use 28 equilateral triangle bin configurations with side lengths spaced in bins between $5\ \text{Mpc}$ and $109\ \text{Mpc}$. These bins are spaced linearly for radii less than $20 \Mpc$, with logarithmically spaced bins for higher radii\footnote{The radius values for these 28 bins are: 1, 2, 3, 4, 5, 6, 7, 8, 9, 10, 11, 12, 13, 14, 15, 16, 17, 18, 19, 20, 24, 29, 35, 42, 51, 62, 75, 91, and 109 Mpc.}. \added{Increasing the number of r-vector configurations beyond these equilateral triangles would almost certainly improve our ability to predict the mean bubble size or ionization fraction history. Further work would be needed however to investigate what size and shape of triangle configurations encode the most information about the topology of the EoR.}

\subsubsection{Mean-free-path measurements for \texorpdfstring{$\xhiir$}{xHII(r)}}
\label{sec:measure_mfp}

\begin{figure}
\includegraphics[width=\columnwidth]{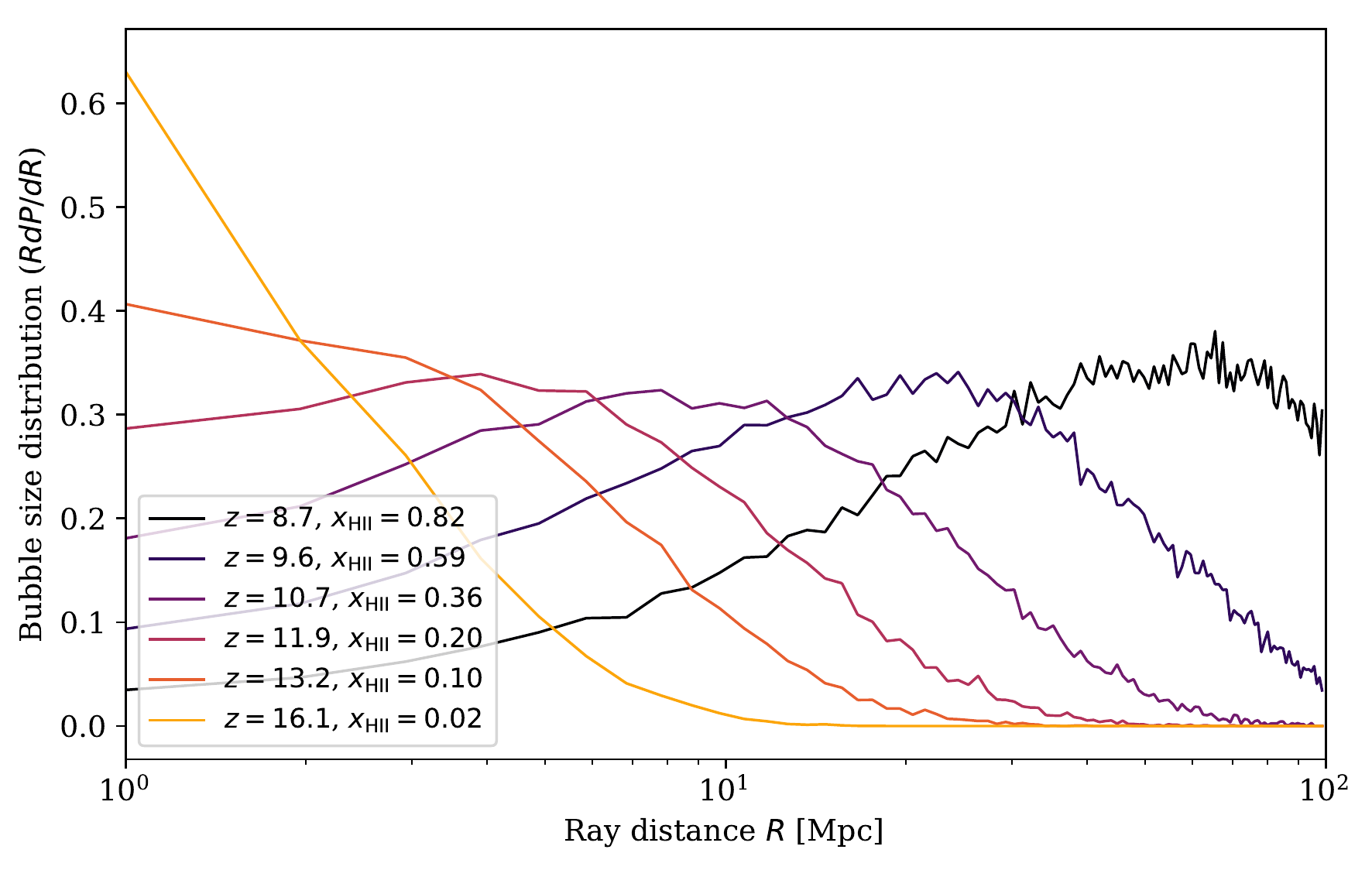}
\caption{Example mean free path measurements of $R dP/dR$ using ionization fraction field data $\xhiir$. Each line shows $R dP/dR$ for a single redshift taken from a simulation with $\Tvir = 10^4 \text{K}$, $\Zion = 30.0$ and $E_0 = 200 \text{ev}$. \removed{For each line the typical bubble size}}
\label{fig:mean_free_path_example}
\end{figure}

In order to measure the \added{mean} bubble size we use our own implementation of the mean-free path method described in \citet{Mesinger2007}. The mean-free path method simulates the emission of photons from random locations within the transparent regions. \added{The distance travelled by each photon before it hits a phase change (from ionized to neutral)} is measured and the resulting number of rays in a range of radius bins is calculated as $dP/dR$. We use $10^5$ simulated photons in our measurements, and the resulting distances are rounded to the nearest pixel size ($L/N = 250.0 \Mpc / 256 = 0.98 \Mpc$). \removedtwo{The distribution of bubble sizes is then directly proportional to $R dP/dR$ (or equivalently $V dP/dV$). }\figref{fig:mean_free_path_example} shows the resulting distributions for $R dP/dR$ from a simulation with canonical parameter values $\Tvir = 10^4 \text{K}$, $\Zion = 30$ and $E_0 = 200\ \text{eV}$. \removed{Giri et al. (2018) note that the peak radius.... }\added{We use the mean of these mean-free path distributions (hereafter written $\Rbubble$) as a statistic to trace the mean bubble size}.

\section{Machine learning techniques}
\label{sec:ML}

The three-point correlation function of 21cm data likely encodes much information about the underlying reionization processes. We can uncover the relationships between $\xip{3}$ and these physical processes by looking at how the 3PCF changes over a range of physical scenarios. We use machine learning techniques to learn these relationships. Our models extract relationships between physical processes and $\xip{3}$ measurements by using simulated data. Our trained models can then use unseen measurements of 3PCF data to make predictions about the physical status of reionization in the data. In particular, we train models that predict the global ionization fraction and the \added{mean} bubble size from $\xip{3}$. Training these models is effectively a form of high-dimensional curve fitting: learning a best-fit functional form $f(\vec{x})$ that maps from a set of input values $\vec{x} = \xip{3}$ to a set of output values ($\Rbubble$ or $\meanxhiiz$). After training, our models can make predictions for new unseen data. For instance, the \added{mean} bubble size model can take measurements of $\xip{3}$ and predict the \added{mean} bubble size. In this section we describe the machine learning techniques we use along with a theoretical description of how they are trained. All models are trained on the same architecture, each on a single node using 16 Xeon E5-2650 cores and 128GB RAM. 700 of our simulations are used for training and validation, and 300 simulations are held back for testing.

\subsection{Artificial neural networks}
\label{sec:mlp_training_theory}
\label{sec:goodness_of_fit}

Artificial neural networks (ANNs) are a common regression technique for learning a complex non-linear relationship between two sets of variables the `inputs' and `outputs'. An ANN represents the relationship in functional form $y_i = f(\vec{x_i})$ by manipulating the inputs $\vec{x_i}$ through a series of weighted summations and function evaluations. For ANNs this series of repeated operations occurs in a series of distinct layers. The values in the first layer $\vec{h^{(0)}}$ are the input variables $\vec{x_i}$. The values from one layer $h^{(l-1)}_j$ affect the values in the following layer $h^{(l)}_j$ according to

\begin{equation}
  \begin{array}{l}
\label{eqn:ANN}
\vec{h^{(l)}} = h^{(l)}_j = \phi_\theta \left(\ \mathlarger{\sum \limits_{i=1}^{N_i}} W^{(l)}_{ij} h^{(l-1)}_{j} \right)\,.
\end{array}
\end{equation}

\noindent The values in each layer are thus a sum over the values in the previous layer, weighted using a set of trainable values $W^{(l)}_{ij}$. The summations into each neuron are passed through an activation function $\phi_\theta(x)$, which determines \removed{whether} the resulting output values that \added{are} passed on to the next layer of neurons. At the end of this process, the final layer contains the network's fitted evaluations of the function, $y_i = f(\vec{x_i})$. Training these models involves choosing the set of weights $W^{(l)}_{ij}$ which most closely mimic the function's behaviour. The `closeness' with which the model mimics the relationship in the training data is quantified using an objective function,

\begin{equation}
\label{eqn:mlp-objective}
\text{Objective} =  \frac{1}{2 N}\sum \limits_{n=1}^{N} \big[ f(\vec{x_n}) - y_n \big]^2  - \frac{\alpha}{2} \sum \limits_{i,j,l} \left(W^{(l)}_{ij}\right)^2
\end{equation}

\noindent so that training is then done by finding the values $W^{(l)}_{ij}$ which minimize this objective function for some training data $(\vec{x_n}, y_n)$. \added{The regularization parameter $\alpha$ in this equation allows finer control over the complexity of the model. A high value of $\alpha$ encourages the training towards simpler models, with more of the weight values $W^{(l)}_{ij}$ being near zero.} Three common activation functions $\phi_\theta(x)$ are the hyperbolic tangent function $\phi_\theta(x) = \tanh(x)$; the logistic function $\phi_\theta(x) = 1/(1+\exp(-x))$ the rectified linear unit ('relu`) function $\phi_\theta(x) = \text{max}(0, x)$. \added{All three activation functions are used during our hyperparameter search method.}

The weights are initialised randomly and are then updated iteratively in order to improve the objective function. Each iteration is known as a single `epoch'. In each epoch, the weights are updated using the current gradient of the objective. By using this gradient, the weights are moved towards a value that should cause the objective function to improve. Our ANNs use the backpropagation algorithm \cite{Werbos1974}, a common technique for efficiently calculating the gradient of the objective function (see \mcitealt{Rumelhart1986} for a more detailed description of this algorithm). The coarseness with which the weights are updated is controlled by a parameter known as the learning rate. A high learning rate means that the weights are changed with a large magnitude at each step. The learning rate can be set to a constant value for all epochs, but it can also adapt to the current speed of the learning. An adaptive learning rate is usually set to decrease if the objective function plateaus (i.e. begins to fall slowly between epochs). It is common to set an upper limit for the number of epochs allowed. We discuss this and other choices made in our models in \secref{sec:hyper_search} later.

\begin{figure}
\includegraphics[width=\columnwidth]{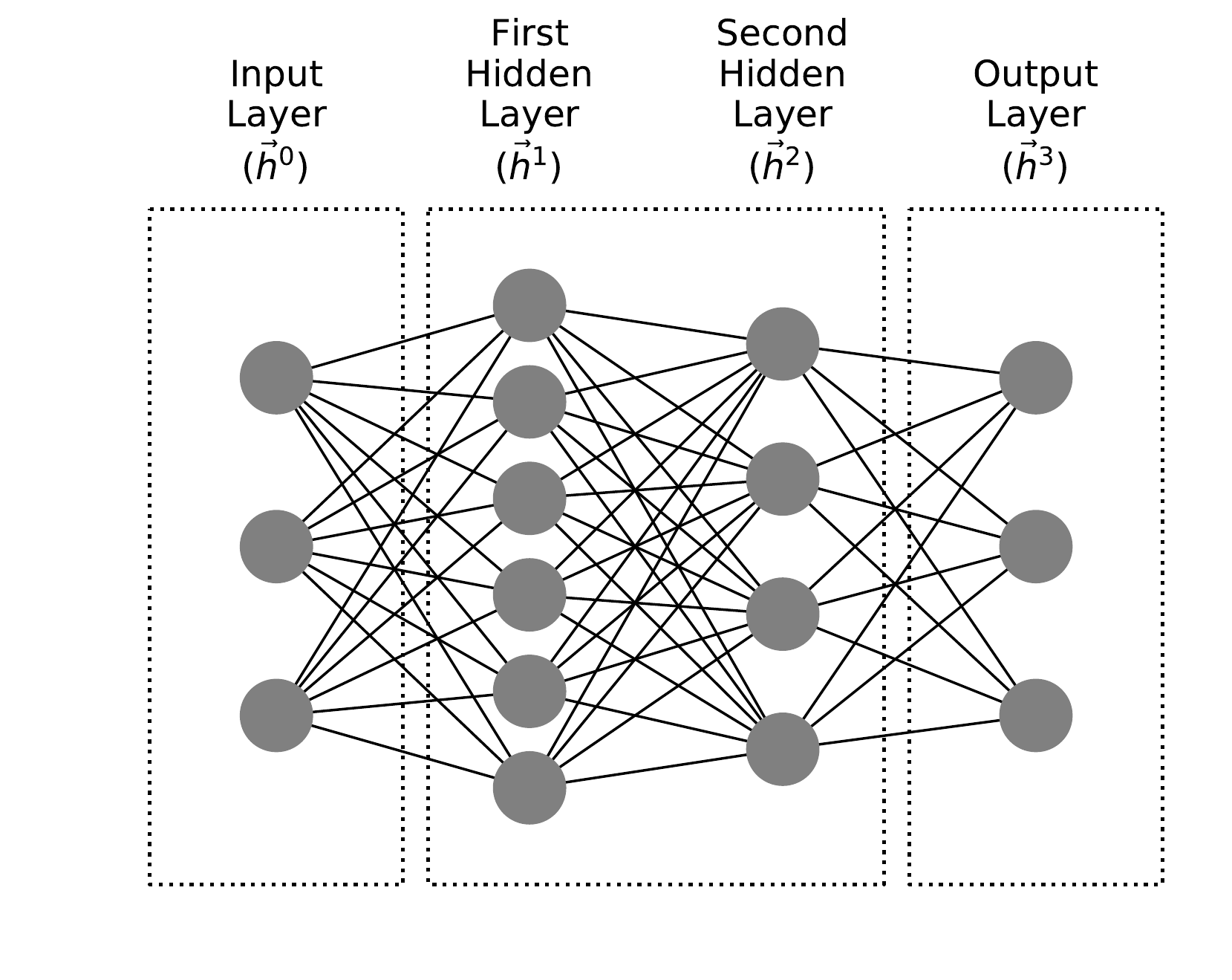}
\caption{Visualization of a multilayer perceptron with two hidden layers. Lines are weighted connections from left to right. Circles are neurons which hold the values and pass them to the following layer.}
\label{fig:mlp}
\end{figure}

Multilayer perceptrons (MLPs) are a subclass of artificial neural networks, with the restrictions that they contain at least one  hidden layer and have a non-linear activation function. \figref{fig:mlp} shows a typical MLP's layer structure, with lines representing the weighted connections between values. Circles represent the neurons which hold the values $h^{(l)}_j$ and pass the weighted inputs through the activation function. Our MLPs are implemented using \code{scikit-learn} from \mcite{Pedregosa2011}. We use the `adam' optimization method \mcitep{Kingma2014} which terminates either when the maximum number of eoochs has been reached, or when the objective function falls below a tolerance of $10^{-10}$ for at least two consecutive iterations.

We measure the goodness of fit between predicted output values $y^*(k,z)$ and measured output values $y(k,z)$ using a mean squared error function,

\begin{equation}
\label{eqn:mse}
\text{MSE} \big[ y, y^* \big] = \frac{1}{N} \sum \limits_{i}^{N}  \left( \frac{y_i - y_i^*}{y_i} \right)^2\
\end{equation}

\noindent also making use of the root mean-squared error $\text{RMSE} = \sqrt{\text{MSE}}$ \added{and the percentage mean-squared error $= 100.0 \times \text{RMSE}$. A percentage RMSE of $100$ indicates that the predicted and measured output values are wrong by an average factor of 2}.

\subsection{Hyperparameter search}
\label{sec:hyper_search}

The weighted-connection values $W_{\mathrm{ij}}$ of a multilayer perceptron are updated during the training process in order to find the best match between the input and outputs in the training data. Several aspects of models must also be fixed before even starting to train the model. We refer to these values as hyperparameters. The hyperparameters can have a strong effect on the final accuracy of predictions but it is rarely obvious what hyperparameter values will result in the most accurate model. We use a random search method with cross-validation to find the best hyperparmeters for each of our model applications. This process is described here.

In order to determine the best hyperparameter values, we train and compare a large number of models with a range of initial hyperparameter values. Each model is trained using a set of randomised hyperparameters and the model with highest prediction accuracy is selected as the best model. Two of the most important hyperparameters are the number of hidden layers and the sizes of these layers, collectively known as the network architecture. The architecture affects the model's ability to represent complex functions: a network with fewer and smaller layers is only be able to model simple relationships, whereas a larger network with more layers (or larger layers) will be able to represent more complex relationships. Using a model that is too small will result in poor prediction accuracy. Using a model that is too large will result in overfitting. There are no prescribed rules for deciding what range of architectures to consider, but a common technique is to use one's knowledge both about the complexity and the dimensionality of the function that is being modelled. When using the 3PCF  measurements as the inputs, there are around $30$ input dimensions to the model. We use networks with between one and three hidden layers, with layer sizes randomly chosen uniformly in the range $[0,500]$. This range of layer sizes was chosen as being a similar order of magnitude to the input dimensionality while also remaining computationally feasible. The full set of parameters which were randomly varied for each model in the hyperparmaeter search are:

\begin{enumerate}
	\item Number of hidden layers uniformly in the linear range $[1, 3]$
	\item Size of each layer uniformly in the linear range $[0, 500]$
	\item Training batch size uniformly in the linear range $[30, 500]$
	\item Number of training epochs uniformly in the range $[50, 500]$
	\item Initial learning rate uniformly in the log range $[10^{-4}, 10^{-2}]$
	\item Learning rate either constant and adaptive with equal chance
	\item Activation from `relu', `tanh', or `logistic' with equal chance
	\item Regularization parameter $\alpha$ from equation \eqnref{eqn:mlp-objective} uniformly in the log range $[10^{-4}, 10^{-2}]$
\end{enumerate}

These ranges match those suggested by the \code{scikit-learn} website \cite{Pedregosa2011}. We use fixed default values for the `adam' parameters $\code{beta\_1} = 0.9$, $\code{beta\_2}=0.999$, $\code{epsilon} = 1e-08$ and $\code{tol} = 0.0001$. For all models the weight values are initialised using the Xavier initialiation strategy \citet{Glorot}. This method sets the weights in the $i$th layer by sampling uniform values in the range $[-U_i, U_i]$. The normalising value $U_i = \sqrt{6} / \sqrt{n_i + n_{i+1}}$ is different for each layer, using values for the total number of input weight connections ($n_i$, also known as `fan in') and output weight connections ($n_{i+1}$, also known as `fan out'). \added{Note that there are seven hyper-parameters being varied in this random search. Given that we choose only 1000 different random sets from the above ranges, it is unlikely that we have identified the\removedtwo{most} optimal model.}

\subsection{Cross-validation}
\label{sec:cross_val}
By trying a range of different hyperparameter values as described we can usually find a model with better prediction accuracy. However this process is sensitive to overfitting. In order to determine which model has the highest accuracy while reducing the chance of overfitting, we use five-fold cross-validation approach. Five models are trained with the same fixed hyperparameters, where each model is provided with data from only four of the five folds. In each case the fifth excluded fold is used to calculate the prediction performance, using \eqnref{eqn:mse}. The performances are thus measured on unseen data, so that the `best' model with highest performance is one which performs well on the unseen validation data. The overall accuracy score is taken as the the mean of the validation scores. \added{This cross-validation approach is used to compare the performance of every combination of hyperparameter values. After finding the best hyperparameter values, the model is trained for a final time using all the training data.} Standard practice for machine learning tasks is to retain a final segment of the data to check the final performance of the best model. If the model performs well on this testing data, then we are more confident that it makes good predictions for completely unseen data. 


\subsection{Input and output scaling}
\label{sec:input_and_ouput_scalings}

Data scaling can be used to improve the efficiency of artificial neural networks during training, and also to improve the quality of the final predictions. The weight values in our neural networks are initialised at small values as described in \secref{sec:hyper_search}. In general, different input features into a model have different scales and magnitudes. Ideally all inputs into the network would have similar orders of magnitude and simple distribution such as normal or uniform. This can easily be achieved by separately normalising or standardising each input feature. Normalising an input feature forces all values to lie in the range $[0, 1]$ by \addedtwo{linearly scaling} the minimum and maximum feature values. \removedtwo{The normalised features}. Standardising an input feature scales the feature to have a mean of zero and a standard deviation of 1.0\removedtwo{, i.e. }. Scaling the model output value(s) also has a beneficial effect on the final prediction accuracy. Our neural networks use the \code{scikit-learn} objective function in \eqnref{eqn:mlp-objective} to quantify the goodness of fit during training. Scaling the output values using normalisation or standardisation can help mitigate the relative importance of output values with different magnitudes.

The input features to our models are the 3PCF measurements $\xip{3}(r)$ for a range of different triangle sizes $r$. These 3PCF values span a wide range of magnitudes. We use the MinMaxScaler method from \code{scikit-learn} to normalise separately each 3PCF bin. We also compare the effect of scaling the 3PCF values by four different powers of the binned radius values: the raw 3PCF $\xip{3}$; the dimensionless 3PCF $r^3 \xip{3}(r)$ used for more natural visualisations (see for instance \citealt{Hoffmann2018}); and two other powers of the radius for completeness: $r \xip{3}(r)$ and $r^2 \xip{3}(r)$. The output features to our models are either the bubble sizes $\Rbubble$ or the global ionization fraction $\meanxhiiz$. We scale the $\Rbubble$ function using the $\text{sinh}^{-1}$ function as described by \citet{Lupton1999}.

\section{Learning typical bubble sizes from the 3PCF}
\label{sec:typical_bubble_size_models}
\label{sec:data_cleaning}

The progress of the Epoch of Reionization can be tracked by measuring the \added{mean} size of ionized regions. Ionized regions are initially small and isolated around the earliest ionising sources. The regions continually grow throughout the EoR, and the precise details of this continued growth depends on the physical interactions between ionising sources and the surrounding neutral regions. The sources themselves are seeded from the clustered non-linear density field and so show significant clustering \mcitep{HultmanKramer2006}, but the details of reionization also affect the clustering of the resulting ionization fraction field $\xhiir$ and 21cm brightness temperature field $\deltatbr$. Throughout the EoR, the \added{mean} bubble size $\Rbubble$ will likely boost the 3PCF at characteristic triangle sizes. Thus, the 3PCF contains information about the physics of reionization \mcitep{Mcquinn2006}. Similarly, higher-order clustering statistics contain information about the physical reionization parameters (see for instance \citealt{Shimabukuro2017a}) which affect the morphology of the $\xhiir$ and $\deltatbr$ fields.

In this section, we train models to predict the \added{mean} bubble size $\Rbubble$ using the 3PCF from simulated data. First, we use 3PCF measurements of the ionization fraction field $\xhiir$ to train our models. The resulting model is a useful means of determining whether $\xip{3}$ does indeed contain information about the \added{mean} bubble size. In practice, however, the ionization fraction field $\xhiir$ is difficult to disentangle from the actual results of interferometer experiments. In the second half of this section we train models to predict the \added{mean} bubble size using simulated $\deltatbr$ data, which would be directly available from interferometer observations. As well as the data cleaning steps in Sections \ref{sec:measure_corr3} and \ref{sec:measure_mfp}, we also exclude data with global ionization fraction outside the range $0.01 \leq \xhii \leq 0.95$.


\subsection{Results training on \texorpdfstring{$\xhiir$}{xHII(r)} data}
\label{sec:corr3_to_bubble_size_xH}
\label{sec:input_scaling_types_results}

\begin{figure}
\includegraphics[width=\columnwidth]{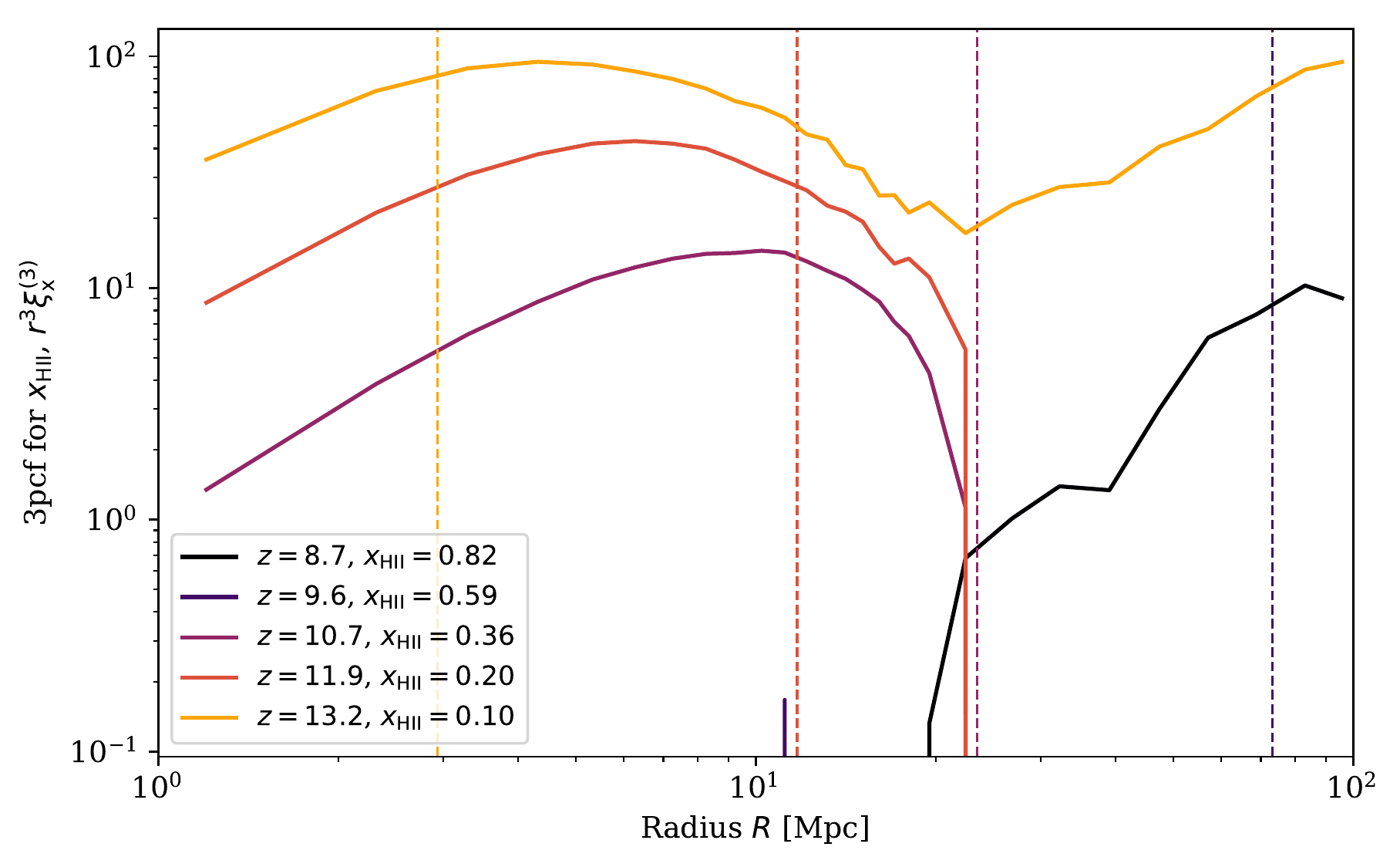}
\caption{Example measurements of $r^3 \xip{3}$ for ionization fraction field data $\xhiir$. Each line shows the measured statistic for a single redshift, all taken from a simulation with $\Zion = 30.0$, $\Tvir = 10^4 \text{K}$ and $E_0 = 200\ \text{eV}$. The redshifts and corresponding global ionization fraction are shown for each line in the legend.}
\label{fig:corr3_xhii_example}
\end{figure}

In this subsection we train a model to learn how the \added{mean} bubble size $\Rbubble$ is related to the 3PCF of ionization fraction data $\xhiir$. Our training and testing data are from the range of simulated reionization scenarios described in \secref{sec:sample_parameter_space}, and we use the multilayer perceptron model described in \secref{sec:ML}. \figref{fig:corr3_xhii_example} shows the measured $\xhii(\vec{r})$ 3PCF for a range of redshifts, showing the true \added{mean} bubble size as vertical lines. This figure is for a scenario with canonical parameter values $\Tvir = 10^4 \text{K}$, $\Zion = 30$ and $E_0 = 200\ \text{eV}$. \removed{The small-scale amplitude of the 3PCF decreases continually, and the amplitude on larger scale increases continually. The turnover radius at intermediate scales also increases throughout the EoR.} \added{The amplitude of the dimensionless 3PCF seen in \figref{fig:corr3_xhii_example} reaches a peak at intermediate scales. Either side of the peak the amplitude decreases, although at larger scales above 20Mpc the amplitude near the start and end of the EoR ($\xhii = 0.82$ and $\xhii = 0.10$) show a second rise in the amplitude.}

Before running a full hyperparameter search, we first compare four options for different input-scaling types. We train one model for each of the four possible input scaling types, namely $\xip{3}$, $r \xip{3}$, $r^2 \xip{3}$ and $r^3 \xip{3}$. The MLP models in this section all have the same architecture, namely two hidden layers both containing $100$ nodes. The following values are used for the other hyperparameters: a training batch size of $200$; $200$ maximum epochs; a constant learning rate of $10^{-3}$; the `relu' activation function; and fixed regularization parameter $\alpha = 10^{-3}$. These hyperparameters were chosen as the midpoints of the allowed random-search ranges or, for categorical choices, as the default parameters suggested by the code authors \citep{Pedregosa2011}. \tabref{tab:rmse_for_r_powers} shows the resulting overall RMSE values for models using each of the four different scaling types. Our results indicate that scaling the 3PCF by $r^2$ or $r^3$ generates more accurate models than scaling by $r$ or not scaling at all. Using $\xip{3}$ or $r \xip{3}$ as inputs makes it harder for our MLP models to uncover a relationship between the 3PCF and the \added{mean} bubble size. We use $r^2 \xip{3}(r)$ as inputs to our models hereafter, as these have the best overall RMSE value. 

\begin{table}
  \centering
  \begin{tabular}{c|c}
    \textbf{Input scaling} & \textbf{RMSE}\\
    \hline
    $\xip{3}$ & \added{6.49} \\
    $r \xip{3}$ & \added{3.59} \\
    $r^2 \xip{3}$ & \added{2.02} \\
    $r^3 \xip{3}$ & \added{2.22} \\
  \end{tabular}
  \caption{RMSE performance on unseen testing data using the four different input scaling types in \secref{sec:input_and_ouput_scalings}. The model using $r^2 \xip{3}$ inputs has the best performance, with the two lowest powers of $r$ having the worst performance. These RMSE values are only for a single cross-validated model with the fixed hyperparameters given in \secref{sec:input_scaling_types_results}, but this indicates that the relationship between $r^2 \xip{3}$ and the \added{mean} bubble size is easier to learn than the other inputs.}
  \label{tab:rmse_for_r_powers}
\end{table}

\subsubsection{Best final model}
\label{sec:best_model_bubble_size_from_r2corr3}

\begin{figure}
\includegraphics[width=\columnwidth]{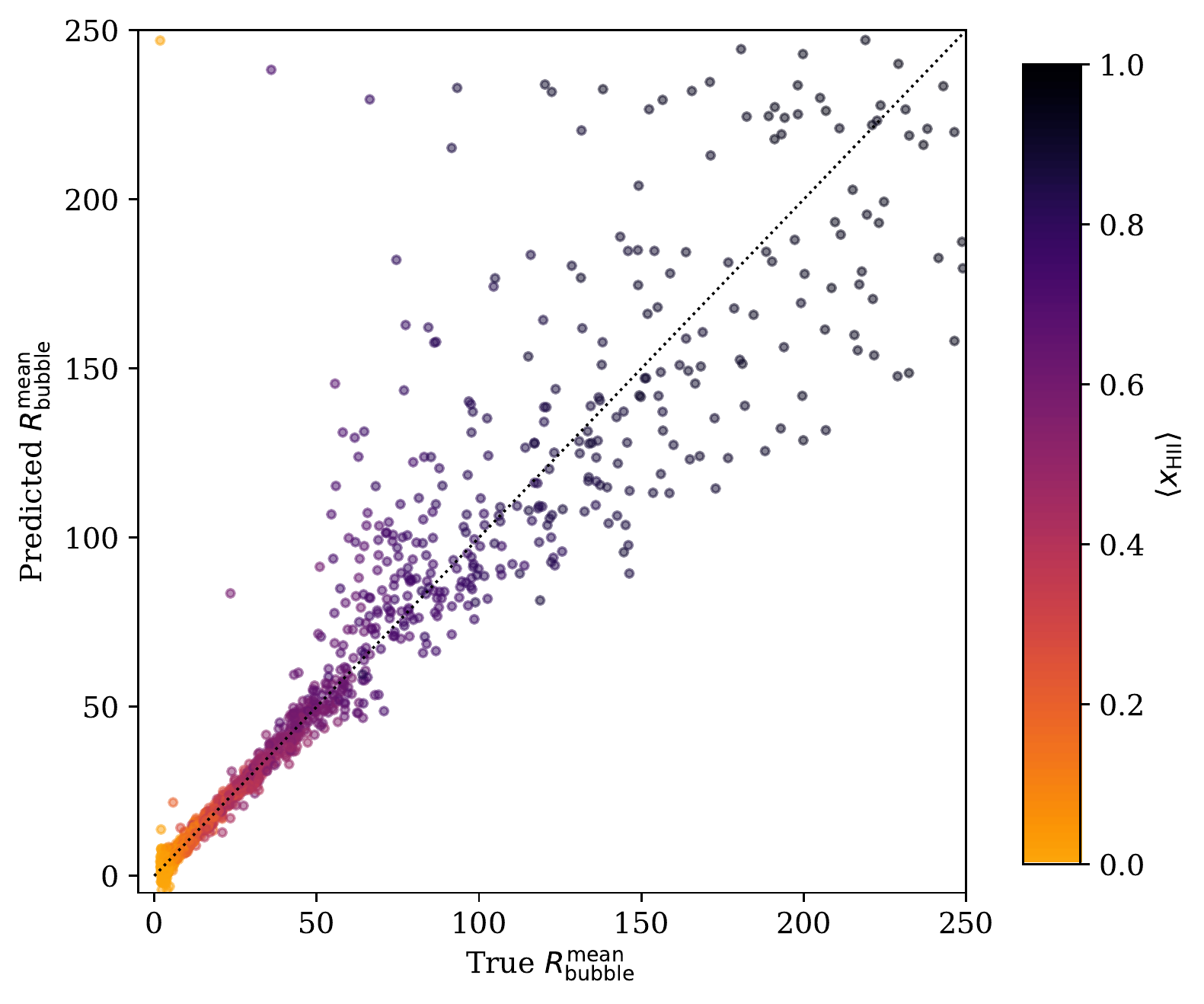}
\caption{Predicted bubble size vs true bubble size for the best model in \secref{sec:best_model_bubble_size_from_r2corr3}. These predictions are made on unseen testing data, using only the 3PCF of ionization fraction field data as inputs to the model. The predicted values and true values generally lie along the diagonal for values of \added{$\Rbubble < 70 \Mpc$}. Larger \added{mean} bubble sizes are harder to model and show much lager scatter away from the diagonal, as discussed in the text.}
\label{fig:corr3_to_bubble_size_xH}
\end{figure}

We now find the best MLP model to predict the \added{mean} bubble sizes from the 3PCF of ionization fraction field data $\xhiir$. We use the full hyper-parameter search method described in \secref{sec:hyper_search}, comparing 1000 randomly chosen models and selecting the one with best cross-validated performance. The resulting best MLP model uses \added{three hidden layers with sizes [148, 142, 93]; training batch size of 296; a maximum of 563 epochs (of which the model used all epochs before terminating); adaptive learning rate starting at $4.7 \times 10^{-3}$; the `relu' activation function; and L2 regularization parameter $3.9 \times 10^{-4}$}. \figref{fig:corr3_to_bubble_size_xH} shows the accuracy of the best MLP model's predictions for unseen testing data. We plot all predicted $\Rbubble$ values as a function of the true values. Marker colours are used to indicate the value of $\meanxhiiz$ for each measurement. A model with perfect predictions would lie exactly on the dotted black diagonal line. Deviations from this diagonal represents less accurate predictions. \figref{fig:corr3_to_bubble_size_histograms} shows the distribution of errors predicted by this model. The median prediction error from these distributions is a good measure of model performance. The model in this subsection has a median prediction error of \added{$10.1\%$}.

The accuracy of the model depends strongly on the magnitude of the true bubble size. \removed{Two interesting features stand out in this figure. First,} The model struggles to make accurate predictions for \added{mean} bubble sizes that are larger than \added{$70\ \Mpc$}: predictions for \added{$\Rbubble < 70 \Mpc$} lie close to the diagonal, but predictions for \added{$\Rbubble > 70 \Mpc$} show much larger scatter. This can be understood in terms of the relationship between the 3PCF and the \added{mean} bubble size. Near the end of the EoR, the widespread overlap of ionized bubbles gives rise to a larger average mean free path of ionising photons, but also blurs the definition of a \added{mean} bubble size. Many bubbles have merged, and thus the `\added{mean}' bubble size is a less clear feature. The model's ability to learn the \added{mean} bubble size from 3PCF measurements reflects this.

\removed{The second feature is the short vertical line}

\subsection{Results training on \texorpdfstring{$\deltatbr$}{deltaTb(r)} data}
\label{sec:corr3_to_bubble_size_delta_T}

\begin{figure}
\includegraphics[width=\columnwidth]{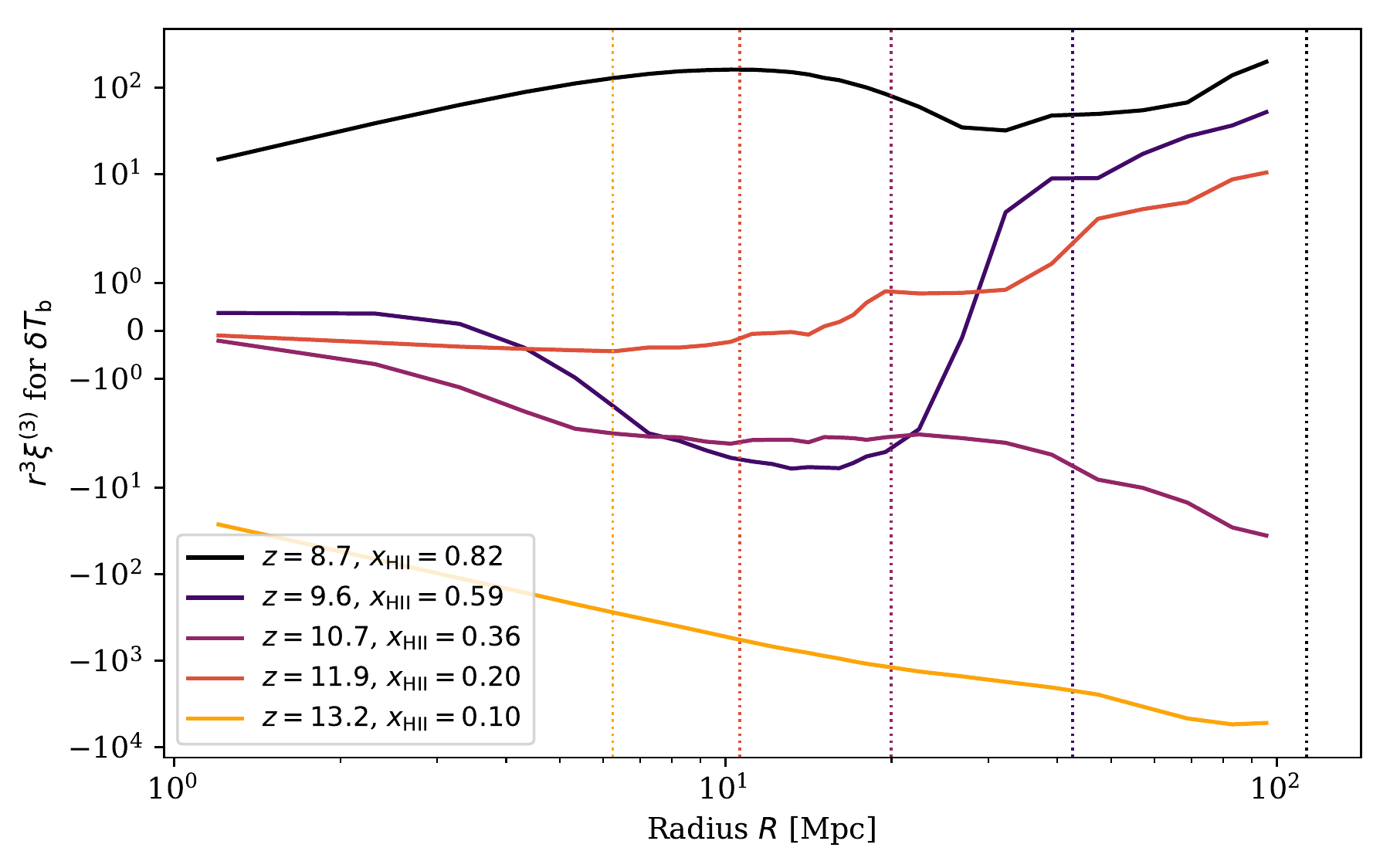}
\caption{Example measurements of $r^3 \xip{3}$ for 21cm differential brightness temperature field data $\deltatbr$, using the same simulation as \figref{fig:corr3_xhii_example}. These data have also been processed using the radius bins in \secref{sec:measure_corr3}.}
\label{fig:corr3_deltatb_example}
\end{figure}

The situation is \removed{considerably} more complicated when using measurements of the 21cm differential brightness temperature field $\deltatbr$ instead of the ionization fraction field $\xhiir$. The relationship between $\deltatb$ and the ionization fraction $\xhii$ given in \eqnref{eqn:deltaTb approx} is assumed to be linear, but the other terms in this equation also impact the morphology of the 21cm brightness temperature field. Most notably, local spin temperature fluctuations $\Tspin(\vec{r})$ and local density fluctuations $\delta(\vec{r})$ can both change the local values of $\deltatbr$. Fluctuations in these values confuse the otherwise simple relationship between the 3PCF and the \added{mean} bubble size. \figref{fig:corr3_deltatb_example} shows the measured $\deltatb(\vec{r})$ 3PCF from a simulation with parameters $\Zion = 30.0$, $\Tvir = 10^4 \text{K}$ and $E_0 = 200\ \text{eV}$. The true \added{mean} bubble sizes \added{(calculated as the mean of the mean-free path distributions)} are shown as vertical lines. The 3PCF of the brightness temperature data has a more complex evolution over the EoR \added{than the 3PCF of ionization fraction data shown in \figref{fig:corr3_xhii_example}}. In general, the amplitude \added{of the $\deltatbr$ 3PCF} decreases until around $\meanxhiiz = 0.25$, before increasing to a maximum near the end of the EoR. The complex evolution of other features is less obvious and justifies the need for machine learning models here.

Using the same method as for the ionization fraction field model, we train a model to predict the \added{mean} bubble sizes using the 3PCF of  simulated $\deltatbr$ data. \added{The resulting best model uses three hidden layers with sizes [158, 188, 187]; training batch size of 169; a maximum of 864 epochs; adaptive learning rate starting at $1.3 \times 10^{-3}$; the `relu' activation function; and L2 regularlization parameter $4.3 \times 10^{-3}$.}

\begin{figure}
\includegraphics[width=\columnwidth]{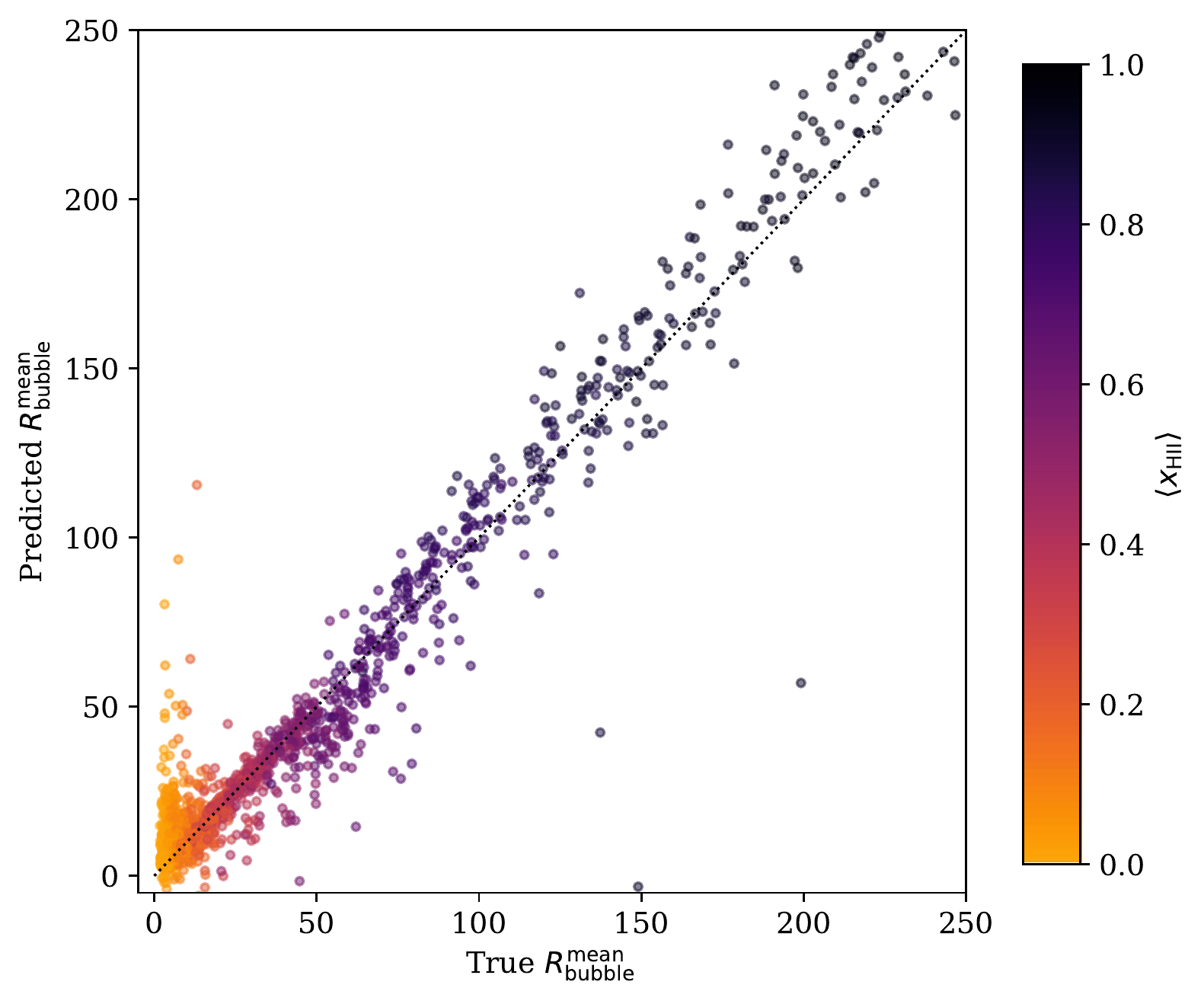}
\caption{Predicted bubble size vs true bubble size for unseen testing data, using the best model in \secref{sec:corr3_to_bubble_size_delta_T}. This model uses the 3PCF of $\deltatbr$ data to predict the \added{mean} bubble size. The predicted values and true values generally lie along the main diagonal for middling values of $\Rbubble$ between 25 and 100 Mpc. The model can accurately predict the \added{mean} bubble size in these scenarios. Deviations from the diagonal line at larger and smaller bubble sizes are worse for the reasons discussed in the text.}
\label{fig:corr3_to_bubble_size_delta_T}
\end{figure}

\added{This $\deltatbr$ MLP model has a median prediction error of $13.4 \%$. This performance is slightly worse than using $\xhiir$ data, indicating that the extra complexities of including local spin temperature fluctuations and local density field fluctuations do indeed contaminate the relationship between the \added{mean} bubble size and the data-field correlations. The model cannot distinguish between correlations of ionized regions and correlations of low density contrast regions (`under-dense' regions), because both of these scenarios give rise to lower values for $\deltatb$. Similarly, regions with low local values for the spin temperature $\Tspin$ can mimic ionized regions.}

\added{We plot the $\deltatbr$ model's predicted \added{mean} bubble sizes for unseen testing data in \figref{fig:corr3_to_bubble_size_delta_T}, as a function of the true \added{mean} bubble size. Two features are worth nothing in comparison to the previous $\xhiir$ MLP model. First, although the average performance of the $\deltatb$ model is worse, the performance at larger bubble sizes is better. Whereas the $\xhiir$ model's predictions showed a large scatter around the diagonal for $\Rbubble > 70 \Mpc$, the $\deltatbr$ model's predictions show a more consistent relationship with the \added{mean} bubble size: all predictions with $\Rbubble > 25 \Mpc$ are made with a roughly consistent accuracy. In particular for larger bubble sizes, the 3PCF of brightness temperature data appears to encode more information about bubble sizes than does the ionization fraction field. A likely reason for this is the effect of neutral regions. Whereas neutral regions in the ionization fraction field have a uniform value of $\xhi = 1.0$, these regions can have different values in the brightness temperature field owing to the other terms in \eqnref{eqn:deltaTb approx}. This relationship could encode information in the brightness temperature field correlations that does not exist in the ionization fraction field, thus allowing our MLP model to learn the mean bubble size more easily. The second interesting feature is the $\deltatbr$ model's poorer performance at low bubble sizes, seen as the large solid cluster of markers in the bottom left of \figref{fig:corr3_to_bubble_size_delta_T}. It is not immediately obvious why this occurs. Including spin temperature fluctuations certainly causes a more complex relationship between the ionization fraction field and the 3PCF of the brightness temperature field. It is possible that this effect is worse at earlier times, when the mean bubble sizes are generally smaller.}

\begin{figure}
\includegraphics[width=\columnwidth]{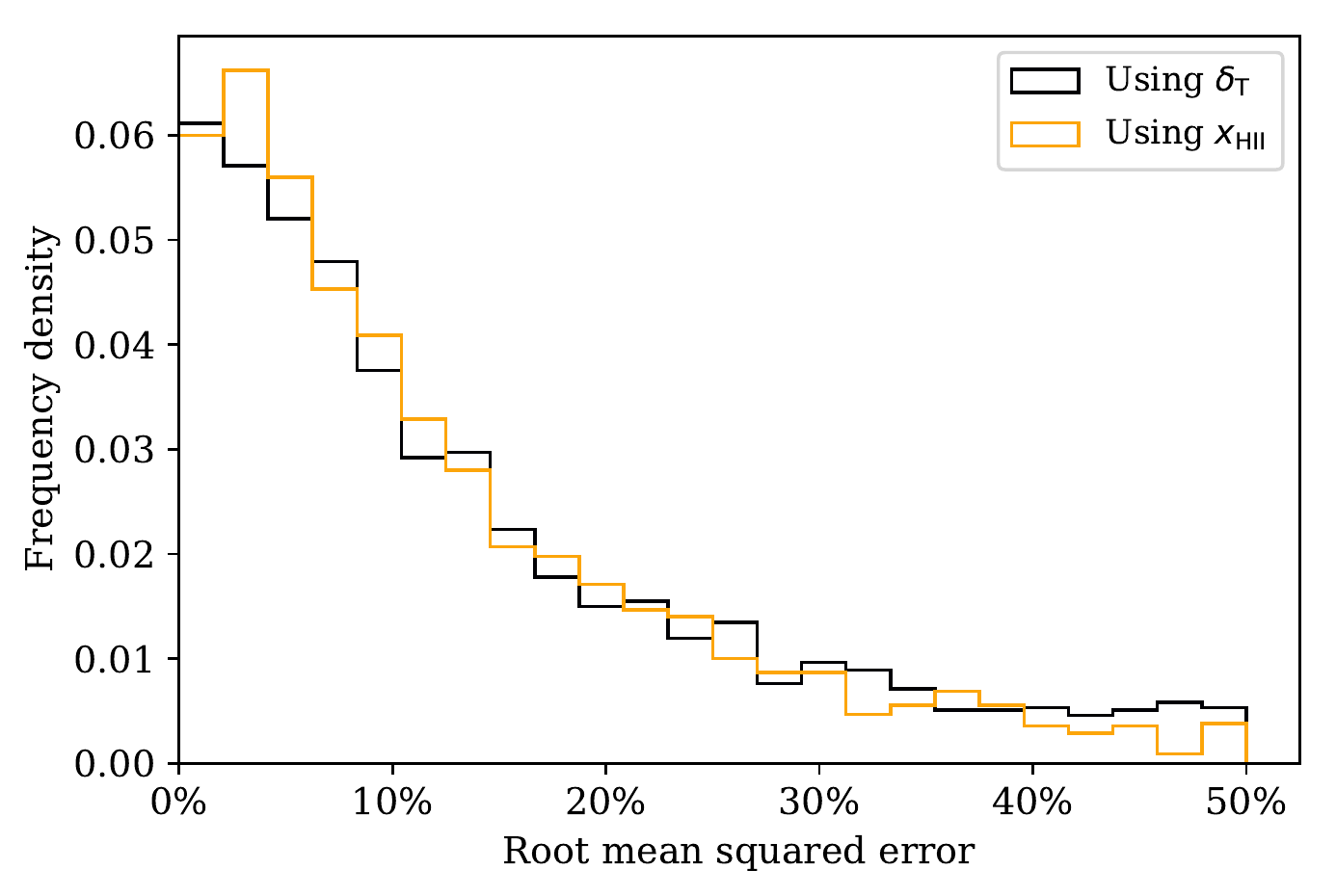}
\caption{Histogram of prediction errors for \added{mean} bubble size models. Visibly, the overall distribution of errors does not depend strongly on which data are used. \added{However, it can be seen from Figures \ref{fig:corr3_to_bubble_size_xH} and \ref{fig:corr3_to_bubble_size_delta_T} that the prediction accuracies of these two MLP models depend strongly on the \added{mean} bubble size: the $\deltatbr$ model makes better predictions at higher bubble sizes, and the $\xhiir$ model makes better predictions at lower bubble sizes.}}
\label{fig:corr3_to_bubble_size_histograms}
\end{figure}

\figref{fig:corr3_to_bubble_size_histograms} shows the histograms of prediction errors for both final best MLP models: one using $\xhiir$ data, and one using $\deltatbr$ data. Ideally, all predictions would be near zero percentage RMSE. The distribution of errors for these two models does not depend strongly on which data are used ($\xhiir$ or $\deltatbr$) although, as mentioned above, each model \removed{does} has different prediction accuracies for different \added{mean} bubble size regimes.

\subsection{Effect of modelling weights}
\begin{figure}
\includegraphics[width=\columnwidth]{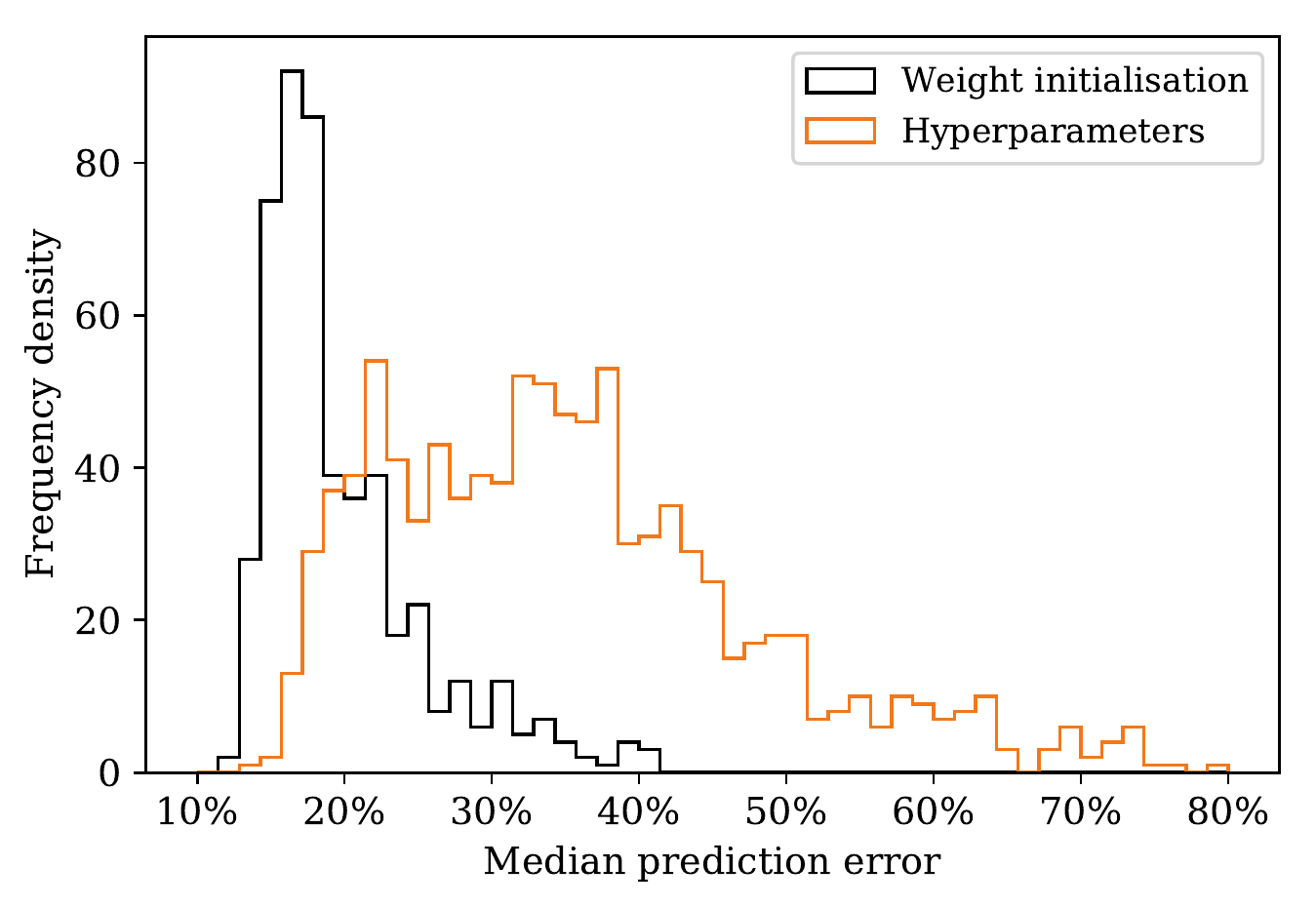}
\caption{\added{Histogram showing the spread of RMSE model performance, either for different weight initialisations of the MLP models or from all 1000 models in the hyperparameter. For the `weight initialisations' line, all models have the best hyper-parameters as determined in \secref{sec:corr3_to_bubble_size_delta_T}.}}
\label{fig:spread_due_to_weights_and_hypers}
\end{figure}

\added{Training an MLP model involves finding the optimal `weight' parameters. These weights are usually initialised to random values as discussed in \secref{sec:mlp_training_theory}. Different initial weight values will result in different final weight values at the end of training. Thus the performance of an MLP model depends on the choice of initial weight values. It is interesting to determine the impact that the choice of initial weight values has on model performance. The black line in \figref{fig:spread_due_to_weights_and_hypers} shows the distribution of median prediction errors for a set of 500 models, each of which has different randomised initial weights but identical hyperparameters to our best model in \secref{sec:corr3_to_bubble_size_delta_T}. The lighter orange line shows the distribution of median prediction errors for all 1000 MLP models in the full hyperparameter search. For our best model, weight initialisation clearly has a strong effect on the model performance: the median prediction error can vary between $10 \%$ and $30\%$. Although it is possible that our best model is particularly susceptible to different weight initialisations, it is likely that other MLP models would also have a similar magnitude of spread in performances. This likely puts an upper limit on the possible performance from any MLP model, even if a deeper hyperparameter search were performed.} \addedtwo{Note that the best model's RMSE value of roughly $13\%$ lies close to the best performance found by varying random initial weights. Our best model has almost certainly benefited somewhat from a `lucky' weight initialisation, in the sense that retraining the MLP model with different initial weights would likely lead to a worse RMSE performance. Our model is a good one -- it has an acceptable RMSE performance on unseen testing data -- but a broader investigation into the full hyperparameter space could potentially lead to a higher accuracy model. } 

\subsection{Comparison to Power Spectrum}
\label{sec:pk_to_bubble_size_delta_T}

\added{The 21cm line is subject to many sources of noise. In particular, thermal noise in the raw observed data affects our ability to make inferences from 21cm maps. In order to reduce the effect of noise, statistical quantities such as clustering statistics can be used. These metrics are less affected by noise since they are calculated as averages across the entire map. The 3PCF is a higher-order clustering statistic and so should reduce the effect of noise. However, it is also interesting to check whether using a lower-order statistic such as the power spectrum provides equally good results. In this section, we use the full hyper-parameter search method described in \secref{sec:hyper_search} to find an MLP model that predicts the \added{mean} bubble sizes from the power spectrum of the differential brightness temperature field $\deltatbr$. As in the previous sections, we compare 1000 randomly chosen models and select the one with best cross-validated performance. The resulting best MLP model uses three hidden layers with sizes [191, 110, 76]; training batch size of 194; a maximum of 596 epochs (of which the model used all epochs before terminating); constant learning rate starting at $4.8 \times 10^{-3}$; the `relu' activation function; and L2 regularization parameter $4.5 \times 10^{-4}$.}

\begin{figure}
\includegraphics[width=\columnwidth]{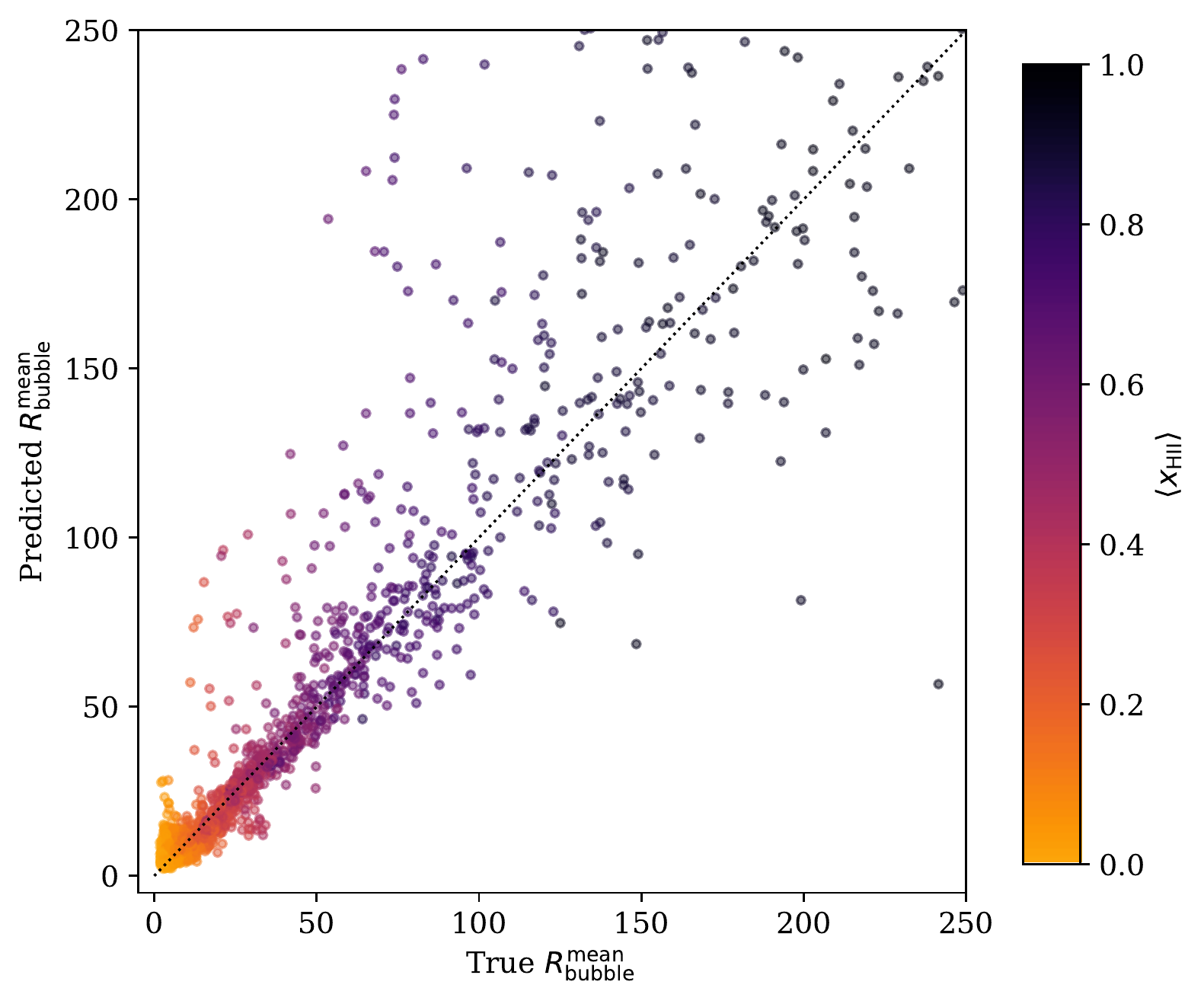}
\caption{\added{Predicted bubble size vs true bubble size for unseen testing data, using the best model in \secref{sec:pk_to_bubble_size_delta_T}. This model uses the power spectrum of $\deltatbr$ data to predict the \added{mean} bubble size. The predicted values and true values generally lie along the main diagonal, although in comparison to the equivalent 3PCF model in \figref{fig:corr3_to_bubble_size_delta_T} this model makes significantly worse predictions.}}
\label{fig:pk_to_bubble_size_delta_T}
\end{figure}

\added{\figref{fig:pk_to_bubble_size_delta_T} shows the accuracy of the best $P(k)$ model's predictions for unseen testing data. This model has a median prediction error of $18.3\%$, somewhat worse than the equivalent model in \secref{sec:corr3_to_bubble_size_delta_T} which uses 3PCF measurements instead of power spectrum measurements as inputs to the MLP model. The information encoded in the power spectrum appears to be less strongly related to the mean bubble size than the information encoded in the 3PCF. It is worth nothing that if noise was added to the underlying simulated $\deltatb$ maps then the performances of the 3PCF models would likely be impacted. More investigation would be needed to determine whether this impact would be greater for the 3PCF model than for the power spectrum models.}

\section{Learning the global ionization fraction from the 3PCF}
\label{sec:global_ionization_fraction_models}

\added{In the previous section we investigate using 3PCF measurements to predict the mean bubble size}. The \added{mean} bubble size is a useful metric for tracking the growth of ionising regions, but the global ionization fraction $\meanxhiiz$ is a more direct measurement for the overall progress of the Epoch of Reionization. The redshift history of $\meanxhiiz$ can be strongly affected by the reionization parameters: different ionising efficiency $\Zion$ scenarios have a different abundance of ionising photons, which affects the EoR duration; different $\Tvir$ scenarios have different halo mass function distributions, leading to more or fewer ionising sources and also affect the EoR duration. 

In this section, we train a model to predict the value of $\meanxhiiz$ from 3PCF measurements. Our models learn the relationship between the 3PCF and global ionization fraction by using the same simulated data in \secref{sec:typical_bubble_size_models}. Measurements of the 3PCF and \added{bubble size distribution} use the methods described in Sections \ref{sec:measure_corr3} and \ref{sec:measure_mfp} \addedtwo{respectively}. The data are cleaned using the same ionization fraction filters, namely $0.01 \leq \meanxhiiz \leq 0.95$.

\subsection{Results training on \texorpdfstring{$\xhiir$}{xHII(r)} data}

We use the same search strategy as in the previous section. The best model uses three hidden layers with sizes [192, 150, 50]; training batch size of 261; a maximum of 365 epochs (of which the model used all epochs before terminating); adaptive learning rate starting at $2.00 \times 10^{-3}$; the \`relu' activation function; and L2 regularization parameter $3.72 \times 10^{-4}$.

\begin{figure}
\includegraphics[width=\columnwidth]{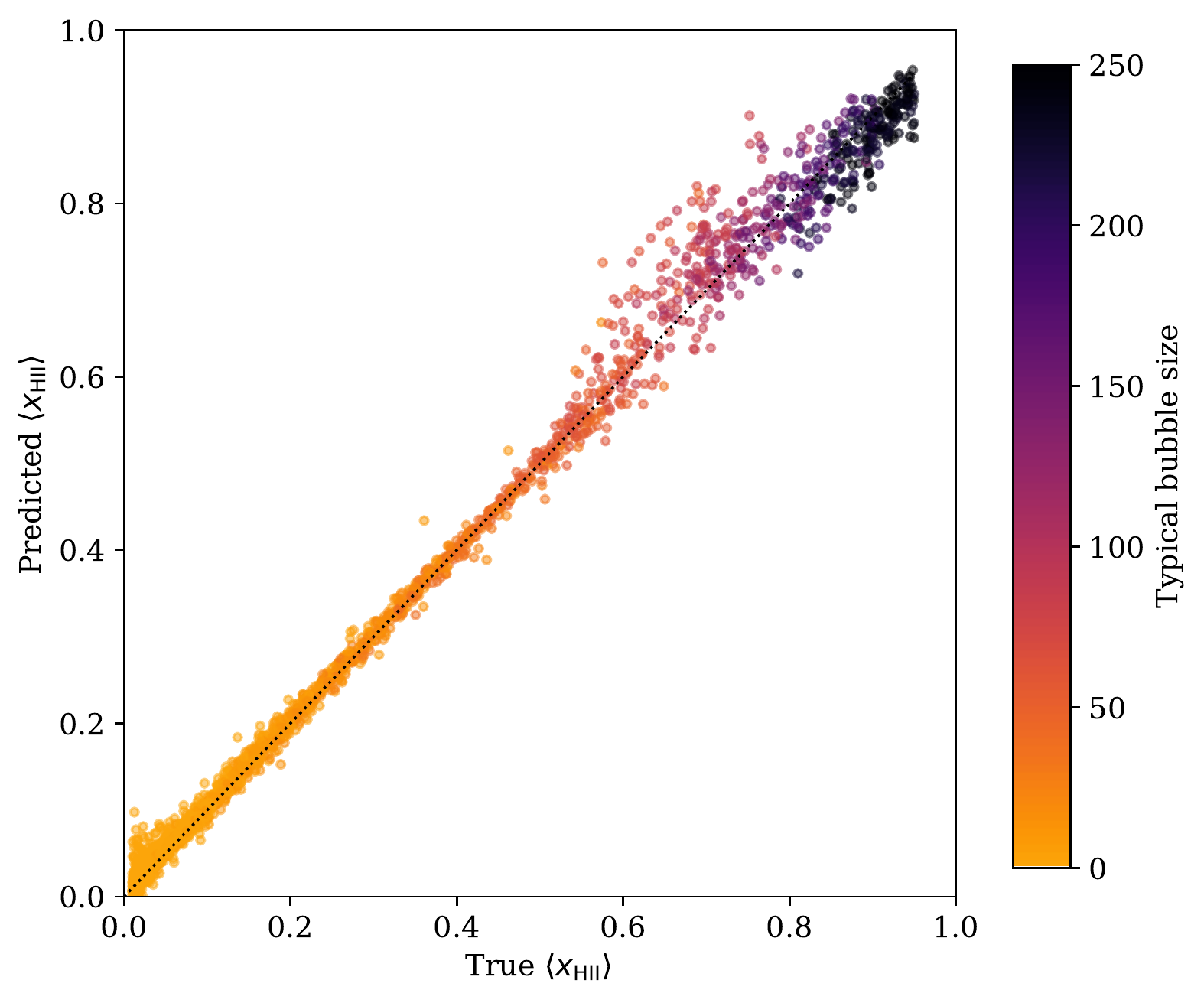}
\caption{Predicted global ionization fraction vs true global ionization fraction for unseen testing data, using the ionization fraction 3PCF as inputs. The predicted and true values lie very closely along the diagonal, particularly for values $\meanxhiiz < 0.6$. Predictions for $\meanxhiiz > 0.6$ are slightly worse as discussed in the text.}
\label{fig:corr3_to_meanxhii_xH}
\end{figure}

The model has an extremely good median prediction error of $3.6 \%$. \figref{fig:corr3_to_meanxhii_xH} indicates the performance from this model, showing the predicted values of $\meanxhiiz$ as a function of the true $\meanxhiiz$ values in the testing data. Marker colours show the \added{mean} bubble size. All markers lie close to the perfect-model diagonal in \figref{fig:corr3_to_meanxhii_xH}, confirming that this model makes extremely accurate predictions. As in the previous section, the model accuracy is higher for $\meanxhiiz < 0.6$ than for $\meanxhiiz > 0.6$.

Ionisation fraction 3PCF measurements have a very strong relationship with the global ionization fraction. Ionisation fraction field data contain a range of bubble sizes. The 3PCF measures clustering on a range of scales and this information is apparently strong enough to provide immediate and accurate predictions for the mean ionization fraction. The predictions begin to worsen near the end of the EoR for $\meanxhiiz > 0.6$, when overlap causes a more complex bubble size distribution. However, the predictions are still visibly good and still have a low RMSE value.

\subsection{Results training on \texorpdfstring{$\deltatb$}{deltaTb} data}
In this subsection we train a model to predict the global ionization fraction $\meanxhiiz$ from $\deltatbr$ 3PCF data. We use the same search strategy as the previous subsections. The best model uses three hidden layers with sizes [168, 174, 70]; training batch size of 361; a maximum of 506 epochs (of which the model used all epochs before terminating); adaptive learning rate starting at $4.44 \times 10^{-3}$; the \`relu' activation function; and L2 regularization parameter $3.65 \times 10^{-3}$. 

\begin{figure}
\includegraphics[width=\columnwidth]{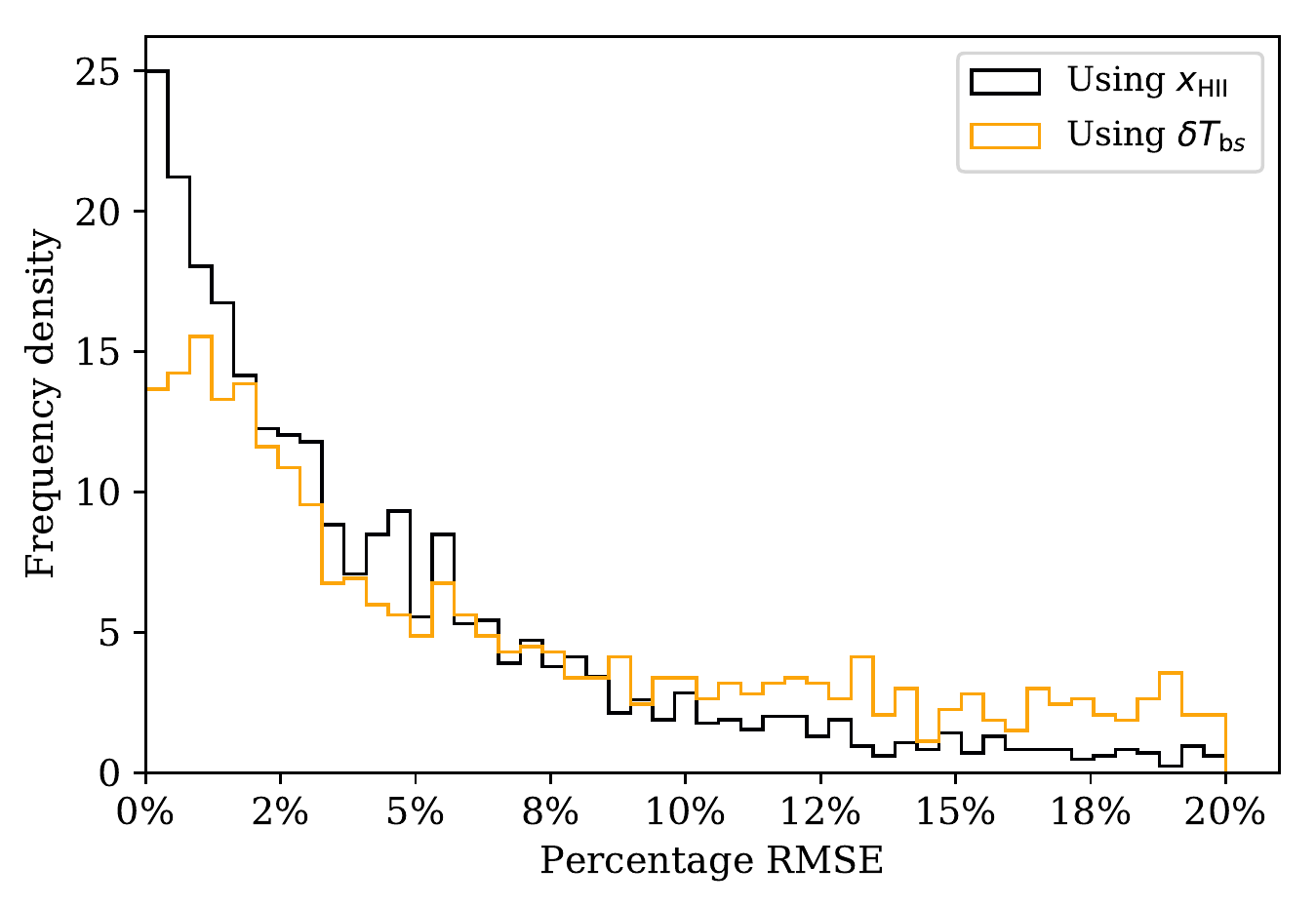}
\caption{Histogram of prediction errors for predicting the global ionization fraction. Each line shows the histogram of errors for a single model. The model using $\xhiir$ 3PCF data has a much more accurate median prediction error ($3.6 \%$) than the model using $\deltatbr$ data ($16.0 \%)$. }
\label{fig:corr3_to_xhii_histograms}
\end{figure}

\begin{figure}
\includegraphics[width=\columnwidth]{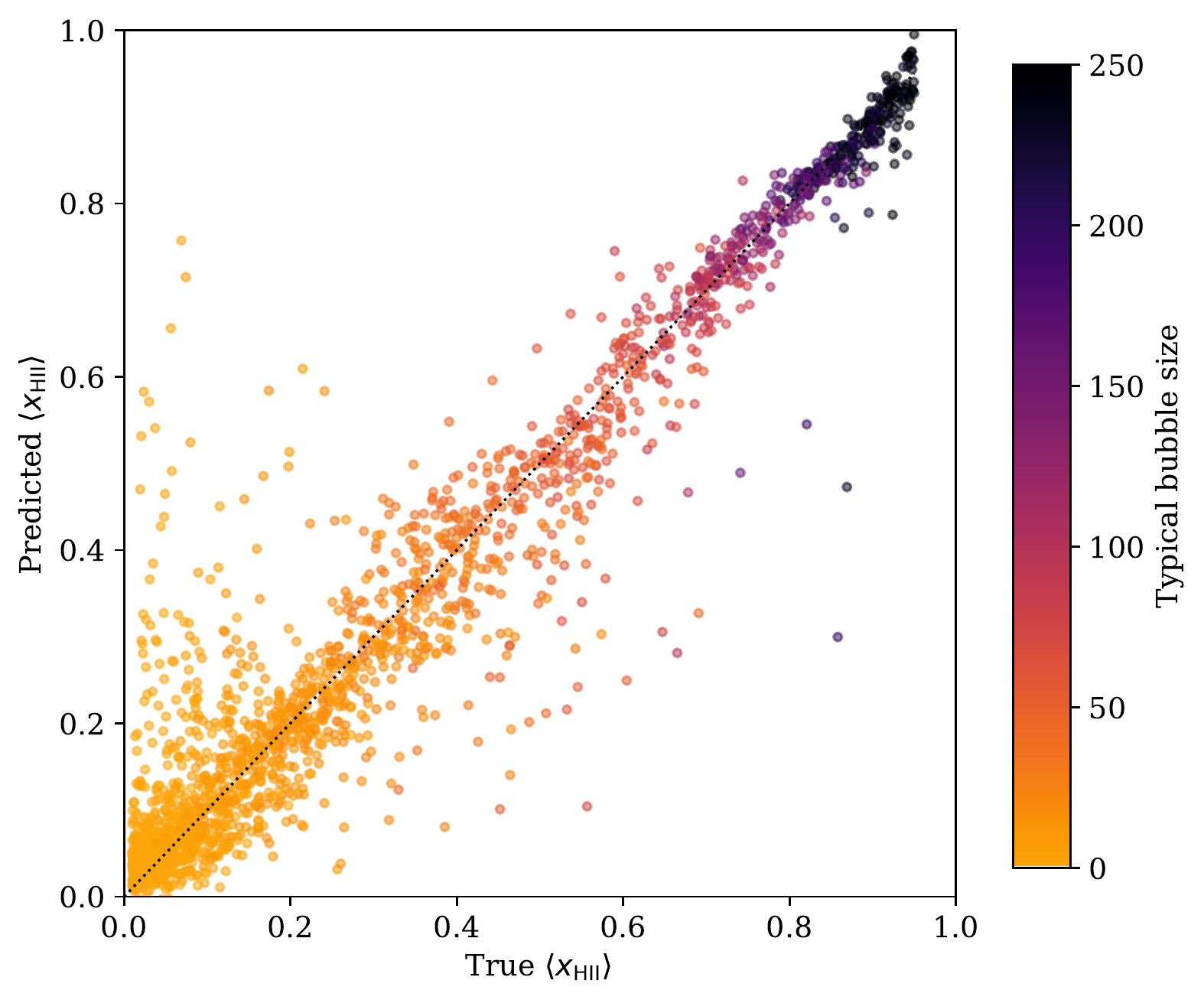}
\caption{Predicted global ionization fraction vs true global ionization fraction for unseen testing data, using $\deltatbr$ as inputs. The predictions generally lie along the diagonal, but with larger scatter than using $\xhiir$ as model inputs.}
\label{fig:corr3_to_meanxhii_delta_T}
\end{figure}

As seen in \figref{fig:corr3_to_meanxhii_delta_T}, predicting the global ionization fraction using $\deltatbr$ 3PCF data gives less accurate results than using $\xhiir$ data. The $\deltatbr$ model's median prediction error is $16.0 \%$, much worse than the error of $3.6 \%$ for the $\xhiir$ model. \figref{fig:corr3_to_xhii_histograms} gives the final prediction histograms for the two global ionization fraction models, using either ionization fraction data $\xhiir$ or brightness temperature field data $\deltatbr$. Predictions of the global ionization fraction depend strongly on which data are used: the prediction errors for the model using $\xhiir$ data are much lower than those for the $\deltatbr$ model.

\removed{Additionally,} The model predictions shown in \figref{fig:corr3_to_meanxhii_delta_T} deviate more widely from the perfect diagonal than do the predictions in \figref{fig:corr3_to_meanxhii_xH}. Interestingly, this model's accuracy \textit{increases} for the later stages of the EoR with $\meanxhiiz > 0.6$, as opposed to decreasing as did the accuracy of the model using $\xhiir$ 3PCF data. This can be understood by considering the impact of density and spin temperature fluctuations. Local fluctuations have a more significant impact on the $\deltatbr$ field at early times than at later times. Thus, the morphology of the $\deltatbr$ field is more closely linked to that of the $\xhiir$ field at later times.

\section{Conclusions}
\label{sec:conclusions}

The three-point correlation function (3PCF) of the 21cm signal contains valuable information about the morphology and history of the Epoch of Reionization. We present an optimised code for estimating the three-point correlation function of 3D pixelised data, such as the outputs from semi-numerical simulations. The code includes jackknifing for error estimates and user-changeable parameters for choosing a level of approximate sampling. We test the code on a testing distribution with known analytic 3PCF, finding that the estimates from our code match the true 3PCF closely. After testing, we use our code to calculate the 3PCF for a range of simulated reionization scenarios using \cmfast{}. \addedagain{Throughout we assume an idealised case where instrumental noise is negligible, and 21cm foregrounds have been perfectly removed.}

We use machine learning techniques and train models to recover both the typical bubble size and the global ionization fraction from measured 3PCF outputs of semi-numerical simulations. We first train models to recover the typical bubble size, from the 3PCF of either ionization fraction data or 21cm differential brightness temperature data. The two models are both able to determine the general trend of increasing typical bubble size and have similar overall accuracy. The model using $\xhiir$ 3PCF data has better performance at small bubble sizes \added{($1 < \Rbubble < 70 \Mpc$}, whereas the model using $\deltatbr$ has better performance for larger bubble sizes \added{($\Rbubble > 25 \Mpc$)}. Both features can be understood in terms of how the data field morphologies evolve over the EoR. \added{We compare the performances of predict the typical bubble size using either the 3PCF or the power spectrum. We find that using the 3PCF instead of the power spectrum leads a noticeable improvement in the final MLP model's prediction accuracy, with median prediction accuracies of around $10\%$ and $14\%$ respectively.}

We then train a model to recover the global ionization fraction from ionization fraction 3PCF data. The resulting model has extremely accurate predictions and shows the three-point clustering of $\xhiir$ data is strongly related to the evolution of the global ionization fraction. Our model is able to uncover this relationship \added{with median prediction accuracy of $4\%$}, although the predictions are slightly less accurate for the later stages of the EoR with $\meanxhiiz > 0.6$. Unfortunately this model would practically not be useful in EoR analysis because the ionization fraction field is difficult to probe directly. Instead, observations are made in terms of the differential brightness temperature. We train a fourth and final model to predict the global ionization fraction from the 3PCF of the differential brightness temperature field. \added{This MLP model has a median prediction accuracy of $16\%$}. The resulting model makes accurate predictions for the late stages of the EoR ($\meanxhiiz > 0.6$), but struggles with the early stages. 

As with all machine learning projects, our models to predict the typical bubble size and global ionization fraction could likely be improved by gathering more data from a wider range of reionizaion scenarios. This would allow the models to learn more general  connections between the 3PCF measurements and characteristic reionization features. Providing other brightness temperature field summary statistics could also improve our models, for instance the distribution of pixel brightnesses \citep{Ichikawa2009} or the size distribution of bright regions \citep{Kakiichi2017}. We also note that our models assume a constant value for the X-ray efficiency. Ideally this constraint should be lifted and the X-ray efficiency allowed to vary as with the other simulation parameters. \added{Further studies will be necessary to evaluate the effectiveness of such an approach in the presence of instrumental effects and noise, as well as foreground residuals.}

\addedagain{The techniques in this paper are tested on simulated data. We have assumed that instrumental noise is negligible at our scales and lower redshifts of operation, as is expected during the EoR upcoming experiments such as the SKA (\citealt{Koopmans2015}). Instrumental smoothing will predominantly affect smaller-scale features on the same scale as the instrument's point spread function, and the effect on larger-scale features would be minimal. \addedtwo{Whilst as noted by \mcitet{Watkinson2018} the bispectrum of Gaussian noise is zero, there will be noise and possibly bias on both the 3PCF and the bispectrum due to sample variance, instrumental systematic effects, ionispheric effects, finite number of baselines, restricted field-of-view, and radio frequency interference. All of these will need to be considered in future studies. Furthermore }we have \removedtwo{also}assumed a best-case scenario where 21cm foregrounds have been perfectly removed. This assumption is not uncommon in recent literature (see for example \citealt{Shimabukuro2017}, \citealt{Gillet2018}, \citealt{Jennings2019}) but remains the subject of much discussion. Several studies (\citealt{Li2019}, \citealt{Mertens2017}, \citealt{Chapman2014}) have claimed that foreground removal can be effective for the power spectrum. \addedtwo{\citet{Watkinson2020b} show that foregrounds could be a problem for recovering the 21cm bispectrum. More work would be needed to understand the impact of foreground residuals on the 3PCF signal.}}

There are several \added{other} possible avenues of future work to build on these results. First, using similar machine learning techniques to predict the full bubble size distribution $dP/dR$ from 3PCF data. The full bubble size distribution provides a more detailed description of the morphology than the typical bubble size alone. \added{Secondly, using a larger selection of triangle configurations (both sizes and shapes) would likely provide more information and make it easier to recover the bubble size statistics.} Thirdly, training models to map from 3PCF measurements directly to parameters in a similar way to \citealt{Shimabukuro2017}. Such inference models can only make estimates of the `best' parameters and do not provide uncertainty regions in the same way as MCMC analysis. Instead, training emulators to forward-model the 3PCF outputs directly from the simulation input parameters would effectively remove the need for further simulations.
\addedagain{Finally, investigating the effect of realistic experiment conditions would indicate whether the 3PCF of future 21cm measurements could be used to extract physically-meaningful bubble size statistics.}

This work presents the first attempt to predict fundamental properties of the Epoch of Reionization using the three-point correlation function and machine learning techniques. We provide a publicly available code \pcffast{} to help the community perform similar analyses in the future.

 \section*{Acknowledgements}
 WDJ was supported by the Science and Technology Facilities Council (ST/M503873/1) and from the European Community through the DEDALE grant (contract no. 665044) within the H2020 Framework Program of the European Commission. CAW's  research  is  supported  by  a  UK  Research  and  Innovation Future  Leaders  Fellowship,  grant  number  MR/S016066/1.  However, the research presented in this paper was carried out with financial support from the European Research Council under ERC grant number 638743-FIRSTDAWN (held by Jonathan Pritchard). FBA acknowledges support from the DEDALE grant, from the UK Science and Technology Research Council (STFC) grant ST/M001334/1, and from STFC grant ST/P003532/1.

\section*{Data Availability Statement}
The data underlying this article will be shared on reasonable request to the corresponding author.

\bibliography{Mendeley}
\bibliographystyle{mnras}

\bsp	
\label{lastpage}
\end{document}